\documentclass[amsmath,amssymb,twocolumn,aps,prx,superscriptaddress]{revtex4-1}

\usepackage{mathtools}
\usepackage{mathrsfs, amsmath, wasysym}
\usepackage[mathscr]{eucal}
\usepackage{xfrac}
\usepackage{graphicx}
\usepackage{dcolumn}
\usepackage{bm}
\usepackage{color}
\usepackage{tabularx}
\usepackage{natbib}
\usepackage[colorlinks,linkcolor=blue,citecolor=blue,urlcolor=blue]{hyperref}
\usepackage[normalem]{ulem}
\usepackage[all]{hypcap}
\usepackage[dvipsnames,usenames,table]{xcolor}

\newcommand{\nocontentsline}[3]{}
\newcommand{\tocless}[2]{\bgroup\let\addcontentsline=\nocontentsline#1{#2}\egroup}

\newcommand{\bk}{{\bf k}}

\newcommand{\bq}{{\bf q}}

\newcommand{\bd}{{\bf d}}

\newcommand{\br}{{\bf r}}

\newcommand{\be}{\begin{equation}}
\newcommand{\ee}{\end{equation}}
\newcommand{\beg}{\begin{gather}}
\newcommand{\eeg}{\end{gather}}
\newcommand{\beq}{\begin{eqnarray}}
\newcommand{\eeq}{\end{eqnarray}}
\newcommand{\bea}{\begin{align}}
\newcommand{\eea}{\end{align}}
\newcommand{\beqq}{\begin{eqnarray*}}
\newcommand{\eeqq}{\end{eqnarray*}}

\newcommand{\ket}[1]{ | #1 \rangle }

\newcommand{\up}{\uparrow}
\newcommand{\down}{\downarrow}
\newcommand{\Up}{\Uparrow}
\newcommand{\Down}{\Downarrow}

\begin{document}

\title{Higher angular momentum band inversions in two dimensions}

\author{J\"orn W. F. Venderbos}
\affiliation{Department of Physics and Astronomy, University of Pennsylvania, Philadelphia, Pennsylvania 19104, USA}%
\affiliation{The Makineni Theoretical Laboratories, Department of Chemistry, University of Pennsylvania, Philadelphia, Pennsylvania 19104, USA}

\author{Yichen Hu}
\affiliation{Department of Physics and Astronomy, University of Pennsylvania, Philadelphia, Pennsylvania 19104, USA}%

\author{C. L. Kane}
\affiliation{Department of Physics and Astronomy, University of Pennsylvania, Philadelphia, Pennsylvania 19104, USA}%

\begin{abstract}
We study a special class of topological phase transitions in two dimensions described by the inversion of bands with relative angular momentum higher than 1. A band inversion of this kind, which is protected by rotation symmetry, separates the trivial insulator from a Chern insulating phase with higher Chern number, and thus generalizes the quantum Hall transition described by a Dirac fermion. Higher angular momentum band inversions are of special interest, as the non-vanishing density of states at the transition can give rise to interesting many-body effects. Here we introduce a series of minimal lattice models which realize higher angular momentum band inversions. We then consider the effect of interactions, focusing on the possibility of electron-hole exciton condensation, which breaks rotational symmetry. An analysis of the excitonic insulator mean field theory further reveals that the ground state of the Chern insulating phase with higher Chern number has the structure of a multicomponent integer quantum Hall state. We conclude by generalizing the notion of higher angular momentum band inversions to the class time-reversal invariant systems, following the scheme of Bernevig-Hughes-Zhang (BHZ). Such band inversions can be viewed as transitions to a topological insulator protected by rotation and inversion symmetry, and provide a promising venue for realizing correlated topological phases such as fractional topological insulators. 
\end{abstract}

\date{\today}


\maketitle



\section{Introduction \label{sec:intro}}

The notion of a band inversion provides a central paradigm for the understanding of free fermion topological phases~\cite{read2000,bernevig2006,fu2007b}. A band inversion marks the transition between two gapped electronic phases in the same symmetry class but with distinct topology, and must necessarily lead to a closing of the energy gap~\cite{hasankane2010,qizhang2011}. 
At the gapless band touching point, where the order of bands is reversed, the topological index associated with the symmetry class changes~\cite{schnyder2008,chiu2016}.
As a result, knowledge of the type of band inversion gives access to information on the topological distinction between the two phases separated by a topological phase transition. This is most clearly exemplified by those band inversions which can be described by a single Dirac fermion theory. In such theories a sign change of the Dirac fermion mass indicates a change of bulk topology. In two dimensions this defines the low-energy theory for the quantum Hall transition~\cite{ludwig1994} and in three dimensions this describes the transition between a trivial and a topological insulator~\cite{fu2007a}. 

In general, when band inversions occur at high-symmetry momenta, the type of such band inversion can be indicated by the eigenvalues of spatial symmetry operators of the bands which invert~\cite{fu2007b,fang2012b,bradlyn2017,po2017,kruthoff2017}. For instance, the Fu-Kane formula can be viewed as a symmetry indicator for a band inversion transition occurring at a time-reversal invariant momentum which changes the $\mathbb{Z}_2$ topological index~\cite{fu2007b}. Another example of established symmetry indicators are crystal rotation symmetries~\cite{hughes2011,fang2012b,fang2012}. Two bands characterized by different crystal rotation eigenvalues have different angular momentum, which implies that, in two dimensions, an inversion of such bands leads to a change of the Chern number (assuming the existence of an energy gap on both sides of the transition). In this paper we study this type of band inversion, with a particular focus on higher angular momentum band inversions. Such band inversions mark the transition to a  Chern insulator with higher Chern number and generalize the transition described by a Dirac fermion.

Our understanding of Chern insulators and Chern bands fundamentally relies on their connection to (flat) Landau levels in a magnetic field~\cite{haldane1988}; as far as their topological classification is concerned, Chern bands and Landau levels are equivalent~\cite{tknn1982}. To a large extent, it is this equivalence, and its implications for properties such as edge state spectrum and Hall conductance quantization~\cite{klitzing1980}, which has motivated and driven much of the research on Chern insulating phases. Furthermore, the connection to Landau levels has been successfully exploited to, for instance, address the effect of electronic interactions in partially filled Chern bands, and thereby explore the possibility of realizing correlated liquid states akin to fractional quantum Hall states without magnetic field~\cite{regnault2011, parameswaran2013, neupert2015, liu2013review}. Here we take a rather different, and in some sense contrary, perspective on Chern insulators, by focusing not on isolated Chern bands but instead on the band inversion transition to the Chern insulating state. Notably, the low-energy description of such transition, which can be viewed as a higher angular momentum generalization of a Dirac fermion transition, exposes a connection to the BCS theory of paired states of fermions in two dimensions~\cite{read2000}. In particular, this connection, which was previously recognized in the context $p+ip$ pairing phases~\cite{liu2011}, suggests that the transition to a Chern insulator phase can be phrased in terms of pairing of electrons and holes---rather than pairs of electrons. One of our aims is to examine this connection in more detail.

We are further motivated by the broader aim to find many-body generalizations of band inversion transitions. In the search for such many-body generalizations higher angular momentum band inversions are of particular interest since the bands disperse quadratically at the critical point of the transition (i.e., when the gap closes and the bands touch, see Fig.~\ref{fig:band-inversion}). This property, which is protected by rotation symmetry, leads to a non-vanishing density of states and implies that---in contrast to band inversion transitions described by a Dirac fermion---interactions are likely to affect the nature of the band inversion~\cite{sun2009,zhang2010,vafek2010,nandkishore2010,uebelacker2011,murray2014,dora2014,hu2018}.  

In previous work~\cite{hu2018} we have argued that, given the importance of interactions, higher angular momentum band inversions provide a promising route towards correlated fluids of electrons and holes. This argument is based on the pairing formulation of the Chern band inversion and was encouraged by the well-established connections between pairing states and fractional quantum Hall wave functions~\cite{read2000}. In this work we focus attention on a second possibility for a correlation-driven phase in the vicinity of the band inversion: the excitonic insulator~\cite{jerome1967}. The excitonic insulator is defined by the condensation of electrons and holes into exciton bound states, which can be called excitonic pairing, and is associated with rotation symmetry breaking.  

To study higher angular momentum band inversions in this paper, we take the following approach. After introducing such band inversions from a low-energy perspective (Sec.~\ref{sec:pairing}), we first construct a class of simple lattice models which realize higher angular momentum band inversions (Sec.~\ref{sec:models}). In this way we take a first step towards identifying material systems in which such band inversions may be observed. We then consider the effect of interactions and address the mean field theory of the excitonic insulator (Sec.~\ref{sec:interactions}). In doing so, we will demonstrate that important insight can be gained into the structure of Chern insulators with higher Chern number~\cite{wang2011,wang2012b,yang2012,trescher2012,kourtis2018} (Sec.~\ref{ssec:GS}). In particular, by studying the transition between the excitonic insulator and the Chern insulator we are able to demonstrate, without making any reference to Landau levels, that the higher Chern number $C=m$ insulator can be viewed as an $m$-component $C=1$ insulator (Sec.~\ref{ssec:m-component}). We conclude by describing time-reversal invariant generalizations of higher angular momentum band inversion (Sec.~\ref{sec:T-invariant}).

\section{Band inversions and Chern insulators \label{sec:pairing}}

\begin{figure}
\includegraphics[width=0.9\columnwidth]{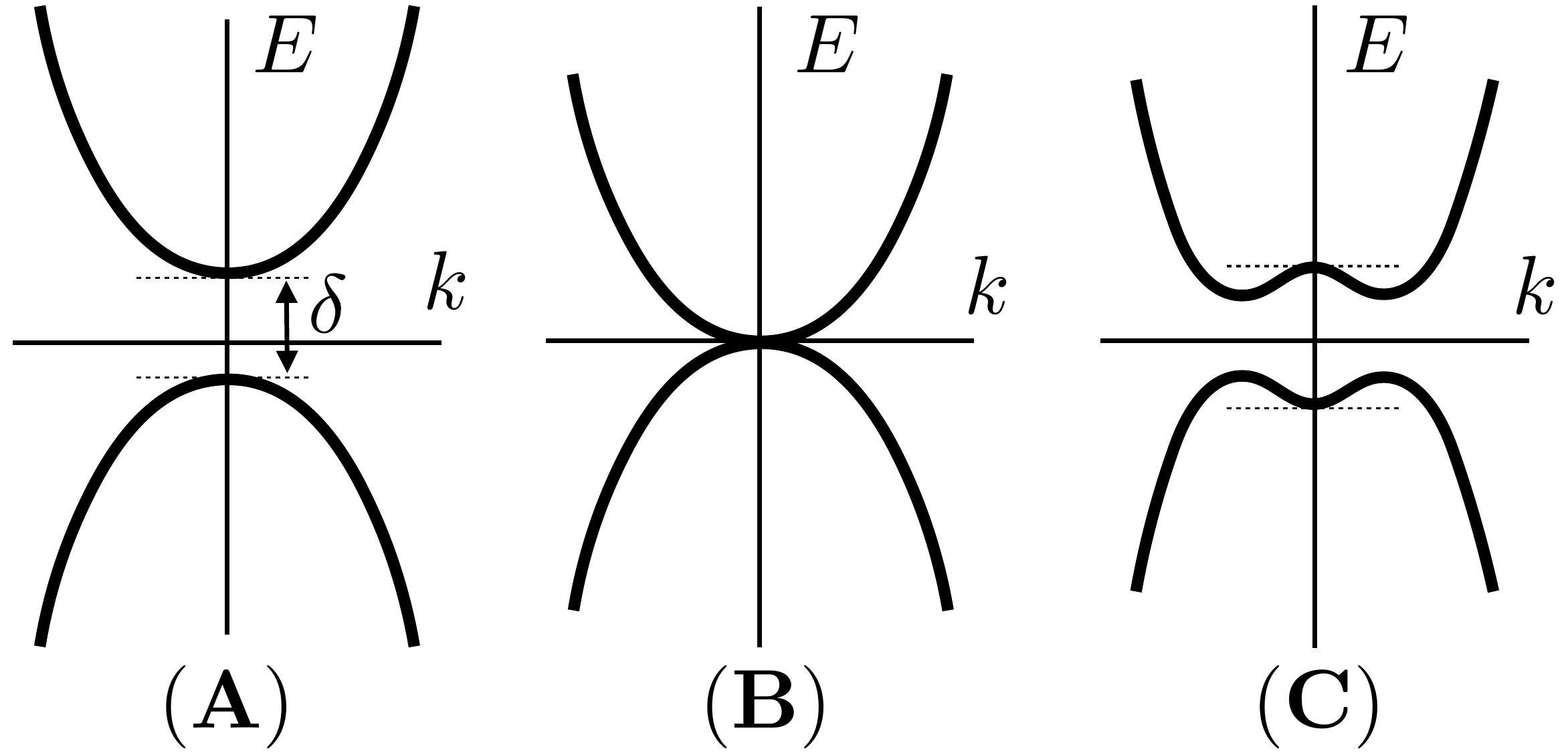}
\caption{\label{fig:band-inversion} {\bf Band inversion transition with higher angular momentum.} A band inversion transition with higher angular momentum in two dimensions separates a trivial insulating phase ($\bf A$) from a topological Chern insulating phase ($\bf C$) with higher Chern number. At the critical point, shown in ($\bf C$), the band dispersion is quadratic, in sharp contrast to band inversion transitions described by a Dirac fermion, for which it is linear. The non-vanishing density of states of the former makes interaction effects relevant, making higher angular momentum band inversions promising venues for many-body generalizations of topological band inversion transitions. 
 }
\end{figure}

We begin by introducing a low-energy theory for band inversion transitions which signal a change of the Chern number index. To describe a band inversion of this type it is sufficient to consider two bands and we thus consider a system with a filled valence band and an empty conduction band, which we study in the vicinity of a band inversion at $\bk=0$. We define the annihilation operators of the conductions band and valance band states as $c_{\bk e}$ and $c_{\bk h}$, respectively, and collect them in the spinor
\be
\psi_\bk = \begin{pmatrix} c_{\bk e} \\  c_{\bk h} \end{pmatrix}.    \label{eq:psi}
\ee
Note that the choice of vacuum (i.e., a filled valence band) implies that $c_{\bk h}$ creates holes in the valence band and can be viewed as a creation operator with respect to the vacuum. In this sense, $\psi_\bk$ may be compared to a Nambu spinor of electrons and holes.  In terms of $\psi_\bk$ and $\psi^\dagger_\bk$ the Hamiltonian can be expressed as 
\be
H= \sum_\bk\psi^\dagger_\bk h_\bk  \psi_\bk , \quad h_\bk=\begin{pmatrix} \varepsilon_\bk & \Delta_\bk \\ \Delta^*_\bk & -\varepsilon_\bk  \end{pmatrix}. \label{eq:H}
\ee
Here $\varepsilon_\bk$ describes the dispersion of the conduction and valence band close to the band inversion at $\bk=0$. To lowest order in momentum the dispersion takes the form $\varepsilon_\bk = \bk^2/2m^* - \delta$, where $m^*$ is an effective mass and $\delta $ is the energy difference between the two bands, which determines whether the bands are inverted ($\delta>0$) or have normal band ordering ($\delta<0$). This is schematically shown in Fig.~\ref{fig:band-inversion}, where ($\bf A$) corresponds to the uninverted regime and ($\bf C$) corresponds to the inverted regime.

It is important to note that $\delta$ is not determined or constrained by symmetry. This should be contrasted with systems exhibiting a symmetry-protected degeneracy of two bands at $\bk=0$, in which case $\delta$ represents a gap opening associated with breaking a symmetry~\cite{sun2009}. Here, on the other hand, we consider a transition between two phases with the same symmetry but different topology. Further observe that the inverted regime $\delta>0$ leads to the notion of an electron-hole Fermi surface defined by the condition $\varepsilon_\bk=0$ and the wave vector $k_F = \sqrt{2m^*\delta}$.
 
The coupling of the electron and hole bands is given by $\Delta_\bk$ and is constrained by the symmetry properties of the electron and hole bands. In this work we focus on a class of band inversion Hamiltonians $h_\bk$ for which the function $\Delta_\bk$ describing the coupling of bands is chiral and characterized by definite nonzero angular momentum $l$. Such couplings with angular momentum $l$ can expressed in the general form
\be
 \Delta_\bk  = \Delta(k_x+ \kappa ik_y)^{|l|} ,\label{eq:Delta-k}
\ee
where $ \Delta$ is the strength of the coupling (which may be complex) and $\kappa= \text{sgn}(l)$. With $\Delta_\bk$ given by Eq.~\eqref{eq:Delta-k} it is straightforward to see that the energy spectrum of $h_\bk$, which consists of two branches $\pm E_\bk$ with $E_\bk  = (\varepsilon^2_\bk + |\Delta_\bk|^2)^{1/2}$, has a full energy gap except for the special case $\delta=0$. This shows that $\delta$ controls the transition between two gapped phases with different topological character, as we now explain.

The form of \eqref{eq:Delta-k} combined with the form of Eq.~\eqref{eq:H} suggests a formal connection to the BCS theory of chiral superconductors in two dimensions~\cite{read2000, nayak2000}. In the latter case, $ \Delta_\bk$ corresponds to the pairing potential and is associated with the breaking of $U(1)$ charge conservation. In this sense, the class of systems we consider here is very different, since all terms present in the Hamiltonian of Eq. \eqref{eq:H}, including $\Delta_\bk$, represent symmetry-allowed couplings between single-particle states. In particular, the number of conduction band electrons and valence band holes is not separately conserved. Given the absence of a broken symmetry one might compare the ``pairing'' of particles and holes described by Eq.~\eqref{eq:H} to proximitized superconductors~\cite{hu2018}. 

The formal connection of Eq. \eqref{eq:H} to chiral superconductors can nevertheless be fruitfully exploited for the purpose of analyzing the ground state wavefunction and its properties. A gapped chiral superconductor in two dimensions with angular momentum $l$ is known to have a topological ground state characterized by a nonzero Chern number $C=l$~\cite{read2000}. This leads to the conclusion that $h_\bk$ with $\Delta_\bk$ given by~\eqref{eq:Delta-k} describes a band inversion transition from a trivial insulator to a Chern insulator with Chern number $C=l$. These two insulating phases are separated by a gap closing at $\delta=0$ (depicted in Fig.~\ref{fig:band-inversion} $\bf B$), with $\delta>0$ corresponding to the Chern insulator, as shown in Fig.~\ref{fig:band-inversion}~($\bf C$).
Following the work of Read and Green~\cite{read2000} the ground state of Eq. \eqref{eq:H} can be expressed in the form
\be
\ket{\Phi} = \prod_\bk (u_\bk  + v_\bk  c^\dagger_{\bk e}  c_{\bk h})\ket{\Omega}  \propto e^{\sum_\bk g_\bk  c^\dagger_{\bk e}  c_{\bk h}}\ket{\Omega},  \label{eq:Phi-GS}
\ee
where $u_\bk$ and $v_\bk$ are solutions to the equations $(\varepsilon_\bk+E_\bk) v_\bk  + \Delta_\bk u_\bk  =0$ and $\Delta^*_\bk v_\bk  +  (E_\bk-\varepsilon_\bk) u_\bk  =0$ with constraint $|u_\bk|^2+|v_\bk|^2=1$, and $\ket{\Omega}$ is the vacuum defined by a filled valence band and empty conduction band (see Appendix \ref{app:pairing}). The ground state $\ket{\Phi}$ describes a Chern insulating phase defined by a ``condensate'' of electrons and holes with nonzero angular momentum $l$. The topology of the many-body wavefunction is encoded in pair correlation function $g(\br) = \int d^2\bk\; g_\bk e^{-i \bk\cdot  \br}/(2\pi)^2$ with $g_\bk = v_\bk/u_\bk$. In Sec.~\ref{sec:interactions} we study the pair correlation function in more detail and discuss its connection to the lattice models introduced in Sec.~\ref{sec:models}.

To address the question how a band inversion of the type defined by Eqs. \eqref{eq:H} and \eqref{eq:Delta-k} can arise, and, in particular, which model systems can describe higher Chern number transitions, it is helpful consider the symmetry properties of $\Delta_\bk$. Since $\Delta_\bk$ is chiral and carries nonzero angular momentum, it can arise when time-reversal and vertical reflection symmetry are both broken. Furthermore, definite angular momentum implies that the form of $\Delta_\bk$ is constrained by rotational symmetry. To see this, consider the case $l=-m$, where $m$ is a positive integer. The Hamiltonian $h_\bk$ can be expressed as 
\be
h_\bk= \varepsilon_\bk \tau_z +\Delta (  k^m_+\tau_-    + k^m_-\tau_+), \label{eq:hk}
\ee
where $\tau_{x,y,z}$ are Pauli matrices and we have defined $\tau_\pm = (\tau_x \pm i \tau_y)/2$ as well as $k_\pm = k_x \pm ik_y$. Under rotations by an angle $\theta$ one has $ k^m_\pm \to  e^{im \theta}k^m_\pm$ and, as a result, one must have $ \tau_\pm \to  e^{im \theta}\tau_\pm$ for $h_\bk$ to be invariant under rotations. We may formulate this in real space by noting that the Hamiltonian takes the form
\be
h= \tau_z(-\partial^2 -\delta) +\Delta \left[ \tau_- ( \partial_{z^*}/i)^m   + \tau_+( \partial_{ z}/i)^m \right], \label{eq:H-z}
\ee
where $\partial_{z,z^*} = \partial_x \mp i \partial_y$. Invariance under rotations implies that the Hamiltonian commutes with the angular momentum operator $L_z$ (i.e., the generator of rotations). To satisfy $[h,L_z]=0$ $L_z$ must have the form
\be
L_z = z\partial_z- z^* \partial_{z^*} + \frac{m}{2} \tau_z , \label{eq:Lz}
\ee
where $z=x+iy$. This leads to the conclusion that the electron and hole bands must have relative angular momentum $m$, i.e., their rotation symmetry quantum numbers must differ by $m$. It this conclusion which provides the basis for the construction of the lattice models in the next section. 

Before we come to a discussion of such models, however, two remarks are in order. First, since the dispersion of the electron and hole band is chosen as $\pm \varepsilon_\bk$,  Eq.~\eqref{eq:H} has a particle-hole symmetry given by $e \leftrightarrow h $ and $l\to -l$. While a convenient starting point for analysis, this is a non-essential assumption and in general one expects this symmetry to be broken by the different band curvature of electron and hole bands. Second, to ensure that the topology of $h_\bk$ is well-defined for $|l|>1$, i.e., that $h_\bk$ is un-inverted at $\bk \to \infty$, higher order terms in $\bk^2$ should be added to $\varepsilon_\bk$.

\section{Lattice models for Chern band inversions \label{sec:models}}

In this section we present a construction of simple lattice models which realize band inversion transitions to Chern insulators with Chern number $C=l$, where $l$ is the angular momentum of the band coupling $\Delta_\bk$ defined in Eq. \eqref{eq:Delta-k}. As demonstrated in the previous section, the constituent degrees of freedom of such models are required to have nonzero relative angular momentum and thus transform nontrivially under the symmetry group of the lattice. Since symmetry plays a central role, we begin by reviewing the generic symmetry properties of Chern insulators and Chern bands and then survey the point symmetry groups compatible with the symmetry requirements of higher angular momentum band inversions. 

Note first that the existence of a Chern insulating state requires broken time-reversal ($T$) and mirror ($M$) symmetry, which follows directly from the transformation property of the Berry curvature under $T$ and $M$ symmetry~\cite{haldane2004}. Here $M$ is a reflection with respect to a vertical mirror plane which inverts one of the coordinates, e.g., $(x,y)\to (x,-y)$. Broken $T$ and $M$ is consistent with the chiral nature of nonzero angular momentum excitonic pairing described by Eq. \eqref{eq:Delta-k}. When the system has multiple inequivalent vertical mirror planes all these reflection symmetries must be broken. As a result, in what follows broken $M$ symmetry should be understood as the absence of all vertical mirror symmetry. A similar result holds for twofold rotations about an axis in the plane, as the Berry curvature is odd under such rotations. 


Chern insulators are compatible with rotation symmetry and our aim is to construct Chern insulator models which preserve the rotation symmetry of the lattice, or, more appropriately, to construct models which exhibit maximal rotation symmetry. The discrete symmetry of the crystal lattice sets limits for rotation invariance: in lattice systems with an $n$-fold rotation symmetry $C_{n=2,3,4,6}$ angular momentum $l$ is only defined $\text{mod}\, n$. As a result, the largest possible angular momentum that can be distinguished is $l=\pm 3$, which implies that the construction of lattice models for excitonic Chern insulators is limited to $C=\pm3$. 
 
In the context of rotationally invariant Chern insulating phases it is worth noting that the relation between the Chern number and angular momentum is also reflected in the fact that the Chern number can be obtained from energy band rotation eigenvalues at rotation invariant momenta (up to multiples of $n$)~\cite{fang2012}.

Next, we examine the crystallographic point groups which may in principle support Chern insulating states with rotation symmetry. Since we consider layer systems with a two-dimensional lattice the appropriate symmetry groups are axial point groups. Admissible symmetry groups are those which leave an angular momentum $l$ along the $z$ axis invariant and allow to distinguish different values of $l$. Consider first the hexagonal groups. There are three groups which satisfy the first condition: $C_6$, $C_{6h}$, and $C_{3h}$. The latter, however, only allows to distinguish $l=\pm1$ and is not of interest. Of the trigonal point groups only $C_3$ and $C_{3i} = S_6$ are compatible with chiral pairing along the $z$ axis. Since $S_6$ includes an inversion $s$-wave and $f$-wave angular momenta have distinct symmetry. In systems with tetragonal symmetry we can only hope to distinguish angular momenta up to $l=\pm2$. Of the groups which preserve angular momentum along $z$, given by $C_4$, $C_{4h}$, and $S_{4}$, all are sufficient to protect $l=\pm2$ pairing. 

To summarize, the symmetry groups of interest are: $C_6$ and $C_{6h}$ (hexagonal); $S_6$ (trigonal); $C_4$, $C_{4h}$, and $S_{4}$ (tetragonal). With this knowledge we now introduce models for systems in these symmetry classes. Our construction is based on the two-dimensional square and triangular lattices with local orbital degrees of freedom at each site.

\begin{figure}
\includegraphics[width=0.6\columnwidth]{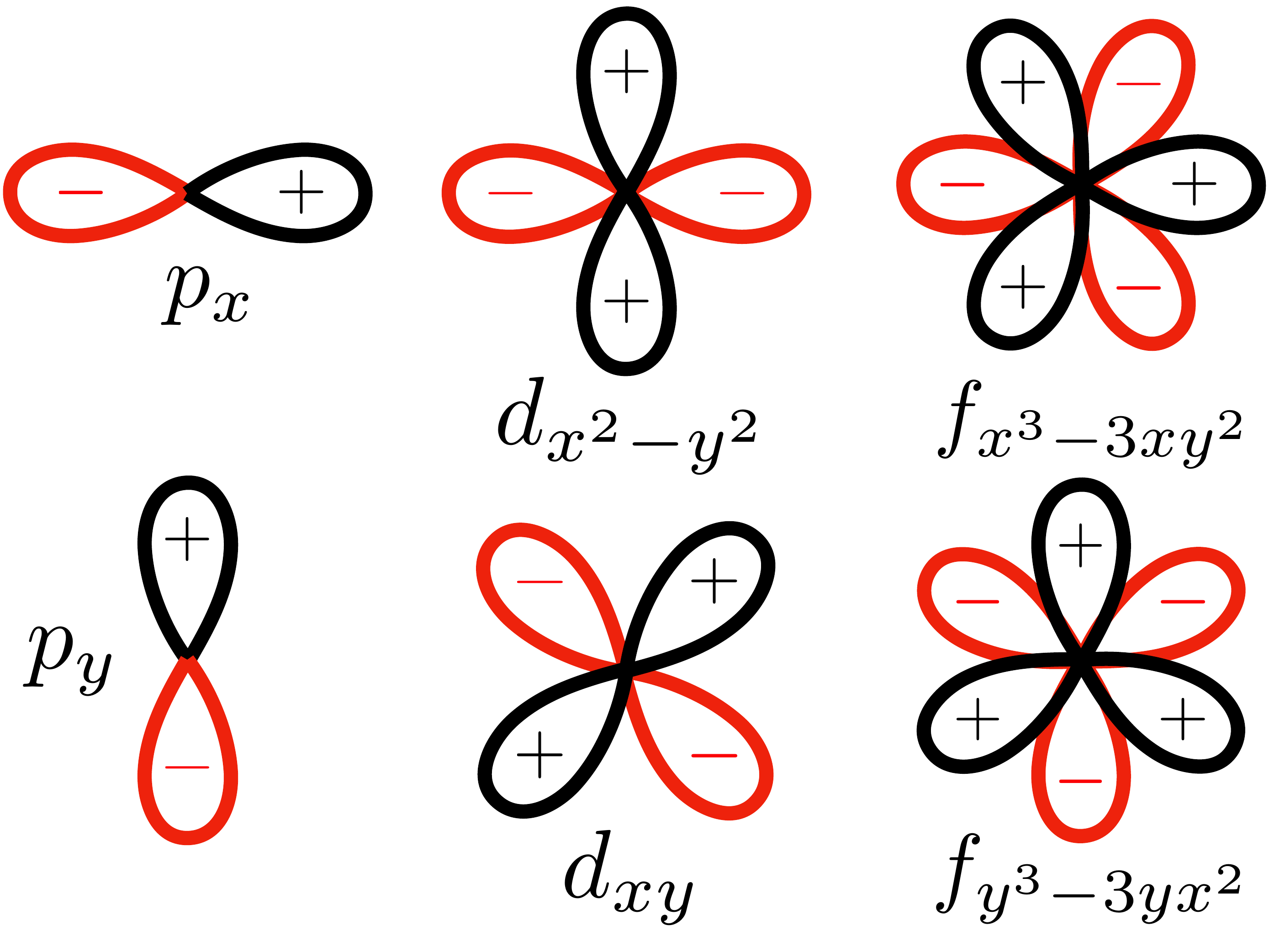}
\caption{\label{fig:orbitals} {\bf Symmetry of orbital states.} Graphical representation of the symmetry of the orbital degrees of freedom with integral angular momentum $l=1,2,3$. The $p$-, $d$-, and $f$-wave states form the basis of the Chern insulator models of Sec.~\ref{ssec:spinless}.
 }
\end{figure}

\subsection{Orbital angular momentum models \label{ssec:spinless}}

\subsubsection{The square lattice \label{sssec:square}}

We first focus on the square lattice. Since the square lattice has $C_4$ rotation symmetry angular momentum can be distinguished up to $l=\pm2$. As a result, the square lattice can support models for band inversion transitions up to Chern number $C=\pm2$. To obtain such models it is natural to choose on-site orbital degrees of freedom with relative angular momentum $\pm2$. We thus consider $s$-wave and $d_{xy}$-wave orbitals and define $s_\bk$ and $d_\bk$ as the electron annihilation operators corresponding to the $s$- and $d$-wave states. (The symmetry of the higher angular momentum orbitals is shown in Fig.~\ref{fig:orbitals}.) We write the Hamiltonian $H$ for this two-band system as
\be
H =   \sum_\bk \varphi^\dagger_\bk h_\bk \varphi_\bk, \quad \varphi_\bk = \begin{pmatrix}  s_\bk  \\   d_\bk  \end{pmatrix},  \label{eq:basisC4}
\ee
where the Hamiltonian matrix $h_\bk$ may be expanded in Pauli matrices $\tau_{x,y,z}$. 

As outlined in the beginning of this section, the form of $h_\bk$ is determined by the symmetry requirements of a $C_4$ symmetric Chern insulator and symmetry of the $s$- and $d$-wave states. An elegant and simple way to derive the form of $h_\bk$ is to formulate the allowed couplings in terms of lattice harmonic functions, which may be viewed as lattice analogs of spherical harmonics and describe hoppings with distinct symmetry. As an example, the (lowest order) $s$-wave harmonic $\lambda^s_\bk$ given by
\be
\lambda^s_\bk =  \cos k_x + \cos k_y,   \label{eq:swaveC4}
\ee
corresponds to the standard nearest neighbor hopping. Note that due to the discrete symmetry of a lattice the lattice harmonics are labeled by the finite set of point group representations (see Table \ref{tab:symmetry}). The two $d$-wave harmonics with $ d_{x^2-y^2} $ and $d_{xy}$ symmetry are given by
\be
\lambda^{d_1}_\bk = \cos k_x - \cos k_y, \quad \lambda^{d_2}_\bk = \sin k_{x}\sin k_{y} ,   \label{eq:dwavesC4}
\ee
and the $p$-wave harmonics are given by $(\lambda^{p_1}_\bk,\lambda^{p_2}_\bk) = (\sin k_{x},\sin k_{y})$.
The symmetry properties and the point group labels of the lattice harmonics are summarized in Table \ref{tab:symmetry} and are shown schematically in Fig.~\ref{fig:orbitals}. 

\begin{figure}
\includegraphics[width=\columnwidth]{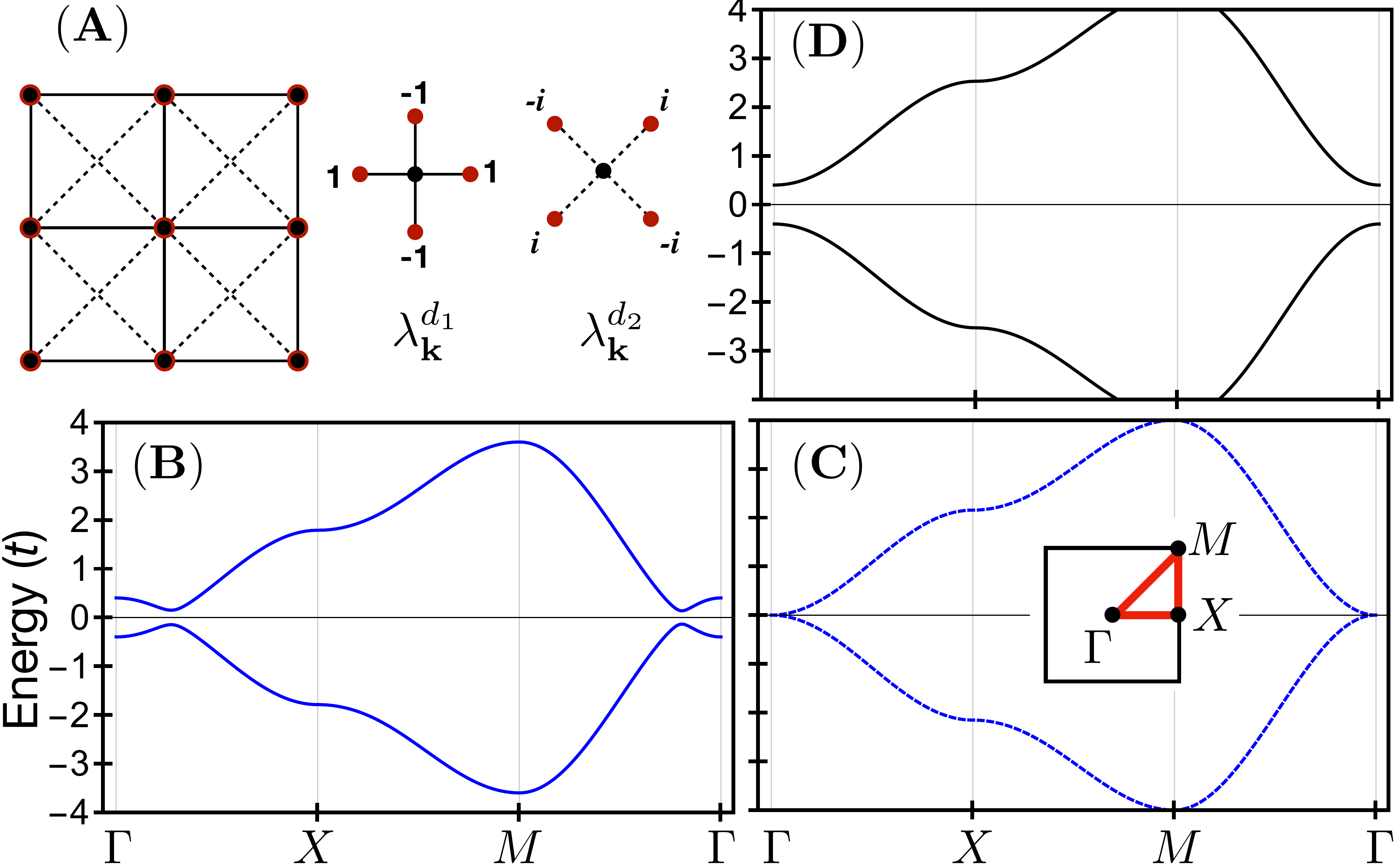}
\caption{\label{fig:square} {\bf Square lattice model.} Panel ($\bf A$) shows the two-orbital square lattice model introduced in Eq.~\eqref{eq:hkC4}, with inter-orbital nearest neighbor hopping ($\Delta_1$) and next-nearest neighbor ($\Delta_2$) hopping. The onsite orbitals with $s$- and $d$-wave symmetry are represented by (superimposed) black and red dots. Shown is also the real space structure of the inter-orbital hoppings defined in Eq.~\eqref{eq:HDelta} and described by the square lattice harmonics $\lambda^{d_1}_\bk$ and $\lambda^{d_1}_\bk$ of Eq.~\eqref{eq:dwavesC4}. ($\bf B$), ($\bf C$), and ($\bf D$) show the spectrum of the square lattice model in the inverted regime, at the critical point, and in the normal regime, respectively.  As parameters we chose $\delta = 0.4t , 0, -0.4t$ and $(\Delta_1,\Delta_2)=(0.4t,0.4t)$.
 }
\end{figure}

Using the symmetry of the orbital basis states and the lattice harmonics it is straightforward to construct a Hamiltonian $H$ which satisfies all symmetry requirements and has a gapped ground state. We write the $H$ as a sum of two parts $H_{\delta}$ and $H_{\Delta}$. Here $H_{\delta}$ describes both nearest neighbor intra-orbital hopping and an energy splitting $\varepsilon_s - \varepsilon_d$ of the $s$ and $d$ states, and $H_{\Delta}$ describes the (inter-orbital) couplings between the $s$ and $d$ states. The splitting between the $s$ and $d$ states is conveniently parametrized as $\varepsilon_s - \varepsilon_d=2t-\delta$, where $t$ is the nearest neighbor hopping parameter; $H_{\delta}$ then takes the form
\be
H_{\delta} =  \sum_\bk (2t-\delta-t\lambda^s_\bk ) (s^\dagger_\bk    s_\bk -d^\dagger_\bk    d_\bk).   \label{eq:Hdelta}
\ee
The structure of $H_{\Delta}$ follows from the observation that $d^\dagger_\bk    s_\bk$ transforms as a $d_{xy}$ wave. The simplest rotationally invariant but $T$- and $M$-broken coupling then takes the form
\be
H_{\Delta} =  \sum_\bk (i\Delta_1\lambda^{d_1}_\bk+\Delta_2\lambda^{d_2}_\bk)  d^\dagger_\bk    s_\bk + \text{H.c.},   \label{eq:HDelta}
\ee
where $\Delta_{1,2}$ are both real and the relative phase is responsible for broken $T$. Combining these two terms we arrive at the form of $h_\bk$ given by
\be
h_\bk  = \varepsilon_\bk\tau_z + \Delta_1\lambda^{d_1}_\bk\tau_y +\Delta_2\lambda^{d_2}_\bk\tau_x,   \label{eq:hkC4}
\ee
where we defined $\varepsilon_\bk=2t-\delta -t\lambda^s_\bk$. The square lattice model defined by \eqref{eq:hkC4} is shown pictorially in Fig.~\ref{fig:square} ($\bf A$). 

It is straightforward to verify that $h_\bk$ has a gapped spectrum for nonzero $(\delta,\Delta_1,\Delta_2)$ and supports Chern bands with $C=\pm2$ for $4t > \delta>0$. The parameter $\delta$ can be directly identified with the band inversion parameter of Eq. \eqref{eq:H}. The spectrum of \eqref{eq:hkC4} is shown in Fig.~\ref{fig:square} ($\bf B$)--($\bf D$), corresponding to the inverted regime ($\delta>0$), the critical point ($\delta=0$), and the normal regime ($\delta<0$).  A more detailed analysis of Eq. \eqref{eq:hkC4} from the perspective of Eq. \eqref{eq:H} will be presented below.

\begin{table}[t]
\centering
\begin{ruledtabular}
\begin{tabular}{lccc}
Symmetry & Lattice & Square  & Hexagonal  \\ 
                & harmonics &  ($D_{4h}$) & ($D_{6h}$) \\ [2pt]
\hline   \\ [-2.0ex]
$s$ & $\lambda^{s}_\bk $ & $A_{1g}$  & $A_{1g}$  \\  [2pt]
$p_x$,  $p_y$ & $\lambda^{p_1}_\bk , \lambda^{p_2}_\bk $  & $E_u$  & $E_{1u}$  \\    [2pt]
$ d_{x^2-y^2} $, $d_{xy}$ & $ \lambda^{d_{1}}_\bk ,  \lambda^{d_2}_\bk $  & $B_{1g}$, $ B_{2g}$   & $E_{2g}$  \\   [2pt]
$f_{x^3-3xy^2}$, $ f_{y^3-3yx^2}$ & $\lambda^{f_1}_\bk,  \lambda^{f_2}_\bk $   & $E_u$  &  $B_{1u}$, $ B_{2u}$ \\  [2pt]
\end{tabular}
\end{ruledtabular}
 \caption{{\bf Symmetry of angular momentum states.} Table summarizing the point group symmetry properties of angular momentum basis functions on the square and hexagonal lattices with (axial) point groups $D_{4h}$ and $D_{6h}$, respectively. These groups are the maximal symmetry groups of a two-dimensional layer. Second column lists the lattice harmonics with given symmetry. Final two columns lists the symmetry quantum numbers. 
 }
\label{tab:symmetry}
\end{table}

\subsubsection{The triangular lattice \label{sssec:square}}

Next, we turn to the triangular lattice, which has sixfold rotation symmetry and allows to resolve angular momentum up to $l=\pm3$. We introduce $s$-wave and $f$-wave states as on-site orbital degree of freedom and define the corresponding electron (annihilation) operator as
\be
 \varphi_\bk = \begin{pmatrix}  s_\bk  \\   f_\bk  \end{pmatrix}.  \label{eq:basisC6}
\ee
As there are two symmetry-distinct $f$ waves, we fix the symmetry by declaring that $f^\dagger_\bk$ creates electrons in a $f_{x^3-3xy^2}$ orbital state, see Fig. \ref{fig:orbitals}. 

To determine the form of the Hamiltonian $h_\bk$ on the triangular lattice we must first specify the triangular lattice harmonics. To this end it is helpful to define the three lattice vectors ${\bf a}_{i=1,2,3}$ as
\be
{\bf a}_i = \begin{pmatrix}  \cos \theta_i  \\   \sin \theta_i  \end{pmatrix} , \quad  
 \theta_i = (i-1)\frac{2\pi}{3}.  \label{eq:triangular}
\ee
The symmetric $s$-wave harmonic then takes the form $\lambda^s_\bk = \sum_{i=1}^3 \cos k_i$, where $k_i= \bk \cdot {\bf a}_i$. The two lowest order symmetry-distinct $f$-wave harmonics are given by
\be
\lambda^{f_1}_\bk = \sum_{i=1}^3 \sin k_i, \quad \lambda^{f_2}_\bk = \frac{1}{3\sqrt{3}}\sum_{i=1}^3 \sin( k_i -k_{i+1}),  \label{eq:C6f-wave}
\ee
where the latter corresponds to next-nearest neighbor coupling (the proportionality constant is chosen for convenience). The $f$ waves $f_1$ and $f_2$ are identified with $f_{x^3-3xy^2}$ and $f_{y^3-3yx^2}$, respectively. In systems with hexagonal symmetry both the $p$ waves $(\lambda^{p_+}_\bk,\lambda^{p_-}_\bk)$ and the $d$ waves $(\lambda^{d_+}_\bk,\lambda^{d_-}_\bk)$ are degenerate, i.e., they form partners of a two-dimensional representation. Expressed in the chiral basis $p_\pm = p_x \pm ip_y$ and $d_\pm=d_{x^2-y^2} \pm id_{xy}$, the triangular lattice $p$- and $d$-waves harmonics take the form
\be
\lambda^{p_+}_\bk = \sum_{i=1}^3 \omega^{i-1}\sin k_i, \quad \lambda^{d_+}_\bk = \sum_{i=1}^3  \omega^{1-i}\cos k_i,   \label{eq:C6pdwaves}
\ee
with $\omega=e^{2\pi i/3}$ and $\lambda^{-}_\bk = (\lambda^{+}_\bk)^*$. Note that the $p_{x,y}$ waves $(\lambda^{p_1}_\bk,\lambda^{p_2}_\bk)$ are simply obtained via the relation $\lambda^{p_\pm}_\bk = \lambda^{p_1}_\bk\pm i \lambda^{p_2}_\bk$, and similarly for the $d$ waves.

\begin{figure}
\includegraphics[width=0.8\columnwidth]{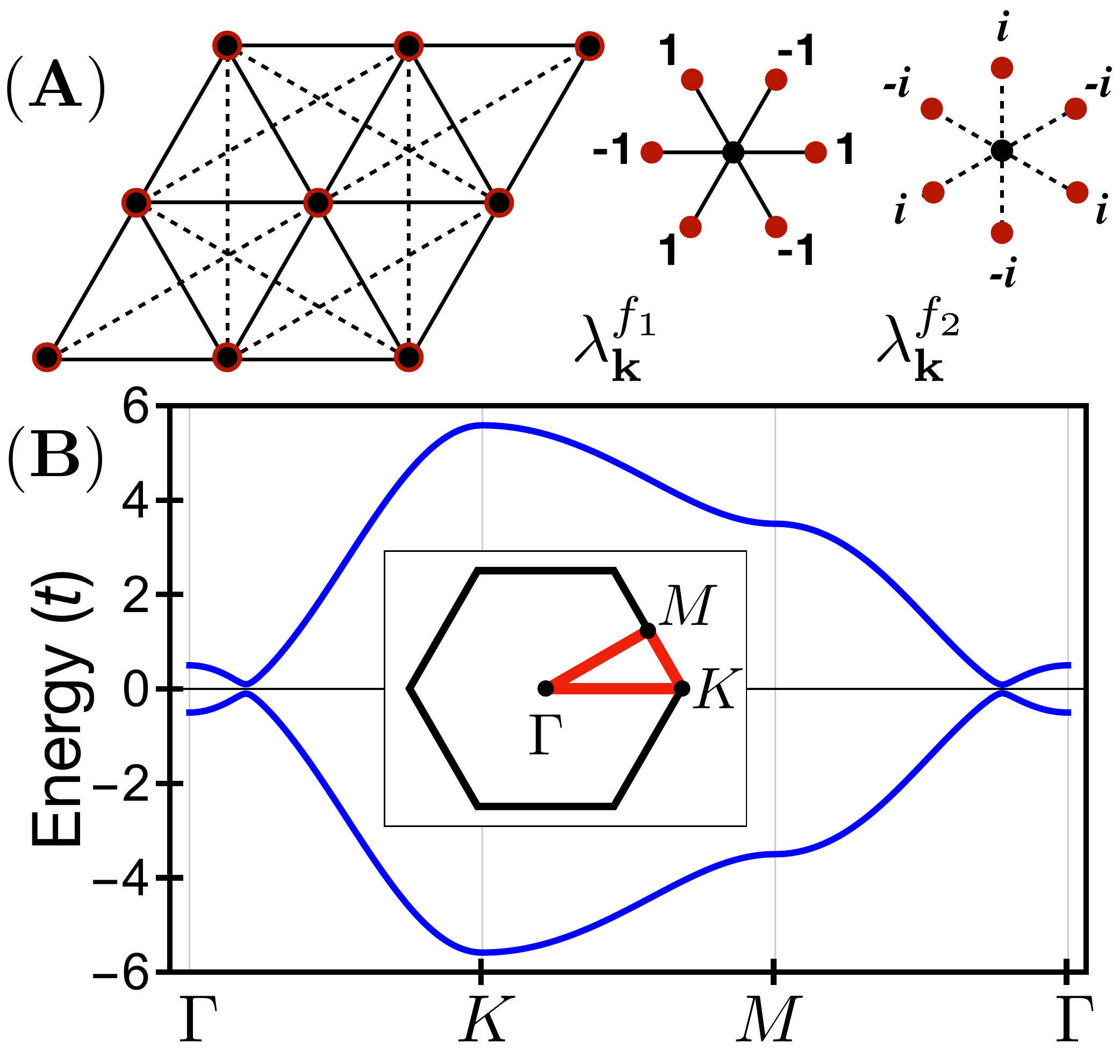}
\caption{\label{fig:triangular} {\bf Triangular lattice model.} 
Panel ($\bf A$) shows two-orbital triangular lattice model introduced in Eq.~\eqref{eq:hkC6}, with inter-orbital nearest neighbor hopping ($\Delta_1$) and next-nearest neighbor hopping ($\Delta_2$). In case of the triangular lattice, the onsite orbitals have $s$- and $f$-wave symmetry, and the real space structure of the inter-orbital hopping, described by the lattice harmonics $\lambda^{f_1}_\bk$ and $\lambda^{f_1}_\bk$, is schematically shown on the right. ($\bf B$) Spectrum of the triangular lattice model in the inverted regime, i.e., $\delta = 0.5t > 0$, for the parameters $(\Delta_1,\Delta_2)=(1.5t,1.5t)$. In inset shows the Brillouin zone path.
 }
\end{figure}

Given the triangular lattice harmonics and their symmetry properties, we directly obtain the triangular lattice analog of Eq. \eqref{eq:hkC4} given by
\be
 h_\bk  = \varepsilon_\bk  \tau_z + \Delta_1\lambda^{f_1}_\bk\tau_y -\Delta_2\lambda^{f_2}_\bk\tau_x.   \label{eq:hkC6}
\ee
Here we have defined the difference of on-site energies $\varepsilon_s - \varepsilon_f$ as $3t-\delta$ and $\varepsilon_\bk = 3t-\delta - t\lambda^s_\bk$, where $t$ denotes ordinary nearest neighbor hopping. The invariance of $h_\bk$ under $C_6$ rotations of follows directly from the symmetry of the $f$-waves couplings. This may be seen, for instance, from Fig. \ref{fig:orbitals}. Since the $f$-wave harmonics are odd function of $\bk$, the second term in Eq. \eqref{eq:hkC6} is invariant under $T$, whereas the third term breaks both $T$ and $M$. A schematic representation of the triangular lattice model of \eqref{eq:hkC6}, in particular the inter-orbital hoppings described by $\Delta_{1,2}$, is shown in Fig.~\ref{fig:triangular} ($\bf A$). For nonzero $(\delta,\Delta_1,\Delta_2)$ the spectrum of $h_\bk$ has a full energy gap and the two non-degenerate bands are Chern bands with $C=\pm3$ when $4t>\delta>0$. A plot of the energy bands in the inverted regime, $\delta=0.5t$, is shown in Fig.~\ref{fig:triangular} ($\bf B$). Note that the gap is proportional to $\delta^{3/2}$.
Below, in Sec.~\ref{sssec:low-energy}, we discuss the low-energy limit of the transition as function of $\delta$ in more detail.

In addition to the triangular lattice model with $C=\pm 3$ bands, it is straightforward to construct a triangular lattice model with $C=\pm 2$ bands. This is achieved by considering $s$-orbital and $d$-orbital states as local degrees of freedom, and since the two $d$-wave states $(d_{x^2-y^2} ,d_{xy})$ are degenerate both should be included a priori. Consider the following model describing the coupling of $s$ and $d$ states, where $d_{\bk 1,2 }$ annihilate electrons with $d_{x^2-y^2,xy} $-orbital symmetry:
\begin{multline}
H = \sum_\bk \varepsilon_\bk(s^\dagger_\bk   s_\bk - d^\dagger_{\bk \alpha}   d_{\bk \alpha})+ \Omega \sum_\bk d^\dagger_{\bk -}   d_{\bk -} \\
+ \sum_\bk \Delta( \lambda^{d_+}_\bk s^\dagger_\bk   d_{\bk -}+\lambda^{d_-}_\bk s^\dagger_\bk   d_{\bk +}) +\text{H.c.}  \label{eq:H-C6-dwave}
\end{multline}
Once more we have defined $\varepsilon_\bk =  3- \lambda^{s}_\bk  +\delta$ and the operators $d_{\bk \pm }  = d_{\bk 1 }\pm i d_{\bk 2 }$ correspond to the chiral basis of the $d$-wave states; a sum over $\alpha=1,2$ is implied. Observe that the term proportional to $\Delta$, which couples the $s$- and $d$-states, is fully invariant under rotations. Furthermore, it is invariant under $T$ and $M$. The second term, on the other hand, which is proportional to $\Omega$ and energetically splits chiral $d$-waves, breaks $T$ and $M$ symmetry. We may choose this energy scale to be positive and very large, i.e., $\Omega \gg 1$, and project out the $d_{\bk -} $ states to obtain an effective model for the $s$ and $d_+$ states. Note that projecting out the $d_{\bk -} $ states is consistent with $C_6$ symmetry and broken $T$ and $M$ symmetry. The reduced two-band model can the expressed in the form of \eqref{eq:basisC4} with $h_\bk$ given by
\be
 h_\bk  = \varepsilon_\bk\tau_z + \Delta  \lambda^{d_-}_\bk\tau_+ +\Delta^*\lambda^{d_+}_\bk\tau_-,   \label{eq:hkC6C2}
\ee
where $\tau_\pm \equiv (\tau_x \pm i \tau_y)/2$. This Hamiltonian describes a transition as function of $\delta$ from a trivial insulator to a Chern insulator with $C=\pm2$ on the triangular lattice. Note that, contrary to Eq. \eqref{eq:hkC6} or \eqref{eq:hkC4}, there is only one parameter $\Delta$ describing the coupling of $s$- and $d$-states, which is due to $C_6$ symmetry.

Clearly, by simply making the replacement $d \to p$ in Eq. \eqref{eq:H-C6-dwave} this construction directly applies to states with $p$-wave symmetry, in which case one obtains a $C=\pm1$ Chern insulator model. Furthermore, the $p$-wave model is easily generalized to the square lattice using the square lattice harmonics \cite{nagaosa2010}, leading to the spinless (and lattice-regularized) Bernevig-Hughes-Zhang (BHZ) model \cite{bernevig2006}.

\subsubsection{The honeycomb lattice \label{sssec:honeycomb}}

Up to this point, we have considered onsite orbital degrees of freedom with nonzero angular momentum. This might suggest that the models introduced here require higher angular momentum atomic-like states (see Fig. \ref{fig:orbitals}) at sites of the crystal lattice. In fact, our construction is more general, and also applies to cases for which effective higher angular momentum states arise as a result of the structure of the unit cell. More specifically, in crystal lattices with a nontrivial unit cell, i.e., a unit cell containing multiple atoms which map to each other under symmetry operations, one can form symmetrized states within the unit cell. These symmetrized states transform nontrivially under the symmetry group in a way that is equivalent to nonzero angular momentum states. Therefore, the orbital states shown in Fig.~\ref{fig:orbitals} should be understood in a more general sense as states of a specific symmetry type, rather than atomic orbitals.  

\begin{figure}
\includegraphics[width=\columnwidth]{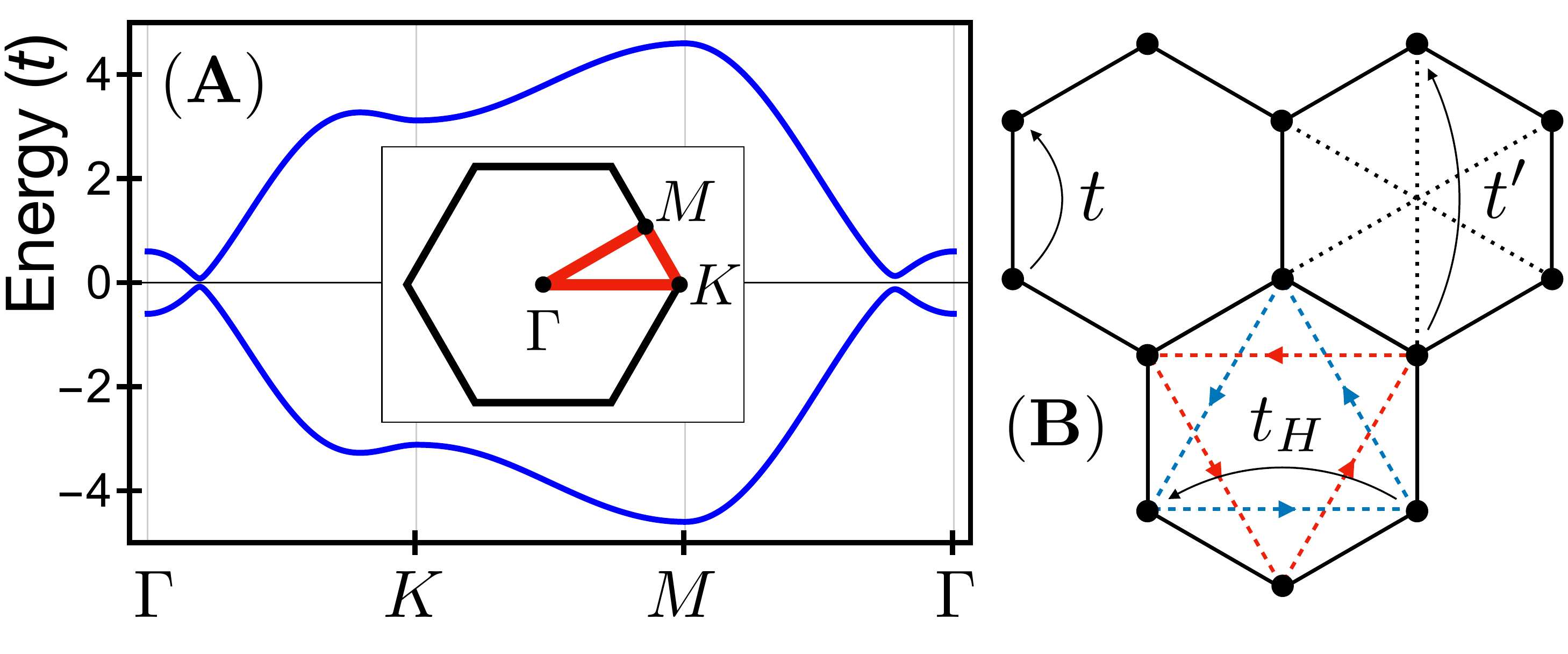}
\caption{\label{fig:honeycomb} {\bf Honeycomb lattice model.} ($\bf A$) Spectrum of the honeycomb lattice model defined in Eq.~\eqref{eq:h-honeycomb} for hopping parameters $(t',t_H)=(1.2t,1.2t)$; the inset shows the Brioullin zone path. ($\bf B$) The honeycomb lattice model is defined by three hopping parameters $t$, $t'$, and $t_H$. Here, $t'$ describes hopping across the hexagon, which is taken to have negative sign in \eqref{eq:h-honeycomb}, and $t_H$ corresponds to the Haldane term and describes $T$-breaking next-nearest neighbor hopping. 
 }
\end{figure}

To illustrate this with an example, we now consider a simple honeycomb lattice model for spinless electrons. The honeycomb lattice, which has a triangular Bravais lattice, consists of two (triangular) sublattices, the $A$ and $B$ sublattice, and we define the corresponding electron operators as $a_{\bk }  $ and $   b_\bk$. As before, we collect these in a spinor
\be
\varphi_\bk = \begin{pmatrix}  a_{\bk }  \\   b_\bk  \end{pmatrix}.  \label{eq:basis-honeycomb}
\ee
The Hamiltonian $H$ is defined as $H = \sum_\bk \varphi^\dagger_\bk h_\bk\varphi_\bk$ with $h_\bk$ given by
\be
h_\bk  = (t\phi_\bk - t'\phi'_\bk)\tau_+  +  (t\phi^*_\bk - t'\phi'^*_\bk)\tau_-  + t_H \lambda^{f_1}_\bk \tau_z.  \label{eq:h-honeycomb}
\ee
Here $\phi_\bk$ is a honeycomb lattice harmonic describing nearest neighbor hopping and is defined as $\phi_\bk = \sum_i e^{i \bk \cdot \bd_i} $, where $\bd_{i=1,2,3}$ are the three nearest neighbor bond vectors $ \bd_{i=1,2,3} = (\sin \theta_i, \cos\theta_i)^T/\sqrt{3}$. [The angles $\theta_{i=1,2,3}$ are the same as in Eq.~\eqref{eq:triangular}.]  Furthermore, the honeycomb lattice harmonic $\phi'_\bk = \sum_i e^{ - 2 i \bk \cdot \bd_i } $ describes third nearest neighbor hopping across a hexagon and the final term proportional to $t_H$ is the Haldane term~\cite{haldane1988}, with $\lambda^{f_1}_\bk$ defined in Eq.~\eqref{eq:C6f-wave}. The three hoppings are shown in Fig.~\ref{fig:honeycomb}~($\bf B$), where arrows indicate $T$-breaking imaginary hopping. 

To see how Eqs.~\eqref{eq:basis-honeycomb} and \eqref{eq:h-honeycomb} give rise to states which have the symmetry of higher angular momentum orbitals consider the Hamiltonian at $\bk=0$. The Hamiltonian takes the form $h_{\bk=0} = (t-t')\tau_x$, which implies that the eigenstates are the even and odd linear combinations $a_{\bk=0 } \pm b_{\bk=0 }$. Clearly, the odd linear combination is odd under all symmetries of the honeycomb lattice which exchange the sublattices and therefore the eigenstates at $\bk=0$ transform as $s$ and $f$ waves. Now, if we redefine $t'=t-\delta$, then $\delta$ parametrizes a band inversion transition of two bands with relative angular momentum $l=3$ at $\bk=0$. As a result, Eq.~\eqref{eq:h-honeycomb} falls in the class of models of which the low-energy description is captures by Eq.~\eqref{eq:H}.

It is easy to recognize that the Hamiltonian of Eq.~\eqref{eq:h-honeycomb} can be viewed as a simple generalization of the Haldane model introduced in Ref.~\onlinecite{haldane1988}. In the context of the Haldane model the band inversion transition described by Eq.~\eqref{eq:h-honeycomb} can be understood as follows. First, we take $t'=0$ but choose $t_H$ nonzero; this is the Haldane model and describes a Chern insulator with $C=\pm1$ bands. Now we turn on and increase $t'$ (which we take positive). As long as the gap stays open the ground state is a Chern insulator with $C=\pm1$ bands. At $t'=t$ the gap closes and reopens for $t'>t$. Since this transition is an angular momentum $l=\pm3$ transition the Chern numbers must have changed by $\pm3$ and indeed we find the resulting bands to have Chern number $C=\mp2$. (Note that the sign of $C$ is determined by the sign of $t_H$.) As a result, neither side of the transition corresponds to the trivial insulator. A plot of the bands for $t'>t$ is shown in Fig.~\ref{fig:honeycomb}~($\bf A$). Note that a large $t_H$ leads to a large separation of bands at $K$, which can be viewed as a large mass for the graphene Dirac points.

\subsubsection{Low-energy limit \label{sssec:low-energy}}

In the models presented above, in particular the square and triangular lattice models, we have made use only of the lowest order lattice harmonics, i.e., we included the nearest (or at most next-nearest) neighbor couplings. As our considerations have shown, for the purpose of constructing models with a low-energy description given by Eqs. \eqref{eq:H} and \eqref{eq:Delta-k} this is sufficient. In general, one may include higher order lattice harmonics in the same symmetry class without affecting the essential physics described by the model.

We now turn to a more detailed analysis of Eqs.~\eqref{eq:hkC4} and \eqref{eq:hkC6} from the viewpoint of higher angular momentum band inversion transitions introduced in Sec.~\ref{sec:pairing}. We begin by expanding the coupling terms of the former, which should be identified with $\Delta_\bk$ of Eq.~\eqref{eq:Delta-k}, to lowest order in $\bk$ and find for $m=2,3$ 
\be
\Delta_\bk \propto (\Delta_1 - \Delta_2)k^m_++(\Delta_1 + \Delta_2)k^m_-,  \label{eq:Delta-expand}
\ee
where $m=2$ and $m=3$ correspond to Eqs.~\eqref{eq:hkC4} and \eqref{eq:hkC6}, respectively. The fact that both $k^m_+$ and $k^m_-$ appear is due to discrete crystal symmetry; the form of Eq. \eqref{eq:Delta-k} is only recovered at a fine-tuned point when $\Delta_1=\Delta_2$. The dominant term is determined by the relative magnitude of $|\Delta_1 + \Delta_2|$ and $|\Delta_1 - \Delta_2|$, which also determines the Chern number in the band inverted regime. By changing one of the two parameters $\Delta_{1,2}$ while keeping the other fixed, the system undergoes a transition from a Chern number $C=\pm m$ phase to a Chern number $C=\mp m$ phase. This transition occurs via a mass inversion at $2m$ Dirac points located on the electron-hole Fermi surface defined by $k_F$ (see Sec. \ref{sec:pairing}). Note that this is consistent with the fact that in a $C_n$ symmetric system the Chern number can only be determined from the rotation eigenvalues mod $n$~\cite{fang2012}; here we have $2m=n$ for $m=2,3$ and $n=4,6$.

Now, let us address the question whether~\eqref{eq:hkC4} and \eqref{eq:hkC6} represent the most general form a Hamiltonian consistent with $C_4$ or $C_6$ rotation symmetry. That is to say, we ask whether there might be additional terms which can be added to~\eqref{eq:hkC4} and \eqref{eq:hkC6} while preserving its generic structure.  In the case of $C_4$ symmetry, we can reconsider Eq.~\eqref{eq:HDelta} and observe that in general $\Delta_1$ and $\Delta_2$ can be complex. This more general Hamiltonian is still symmetric under $C_4$ rotations and translates into an additional term in Eq.~\eqref{eq:hkC4} given by $\Delta'_1\lambda^{d_1}_\bk\tau_x +\Delta'_2\lambda^{d_2}_\bk\tau_y$. Expanding this full $C_4$-symmetric Hamiltonian, with four real parameters describing the coupling, in small momenta $\bk$ one finds
\be
\Delta_\bk \propto \tilde \Delta_+ k^2_++\tilde \Delta_- k^2_- , \label{eq:Delta-k-C4}
\ee
with $ \tilde \Delta_\pm$ given by
\be 
 \tilde \Delta_\pm =  \Delta'_1 \pm \Delta'_2 -i(\Delta_1 \mp \Delta_2) .  \label{eq:Delta+-}
\ee
From this we conclude that a full account of the symmetry-allowed couplings leads a low-energy Hamiltonian of the form Eqs. \eqref{eq:H} and \eqref{eq:Delta-k} with $\Delta_\bk$ given by \eqref{eq:Delta-k-C4} in terms of $\tilde \Delta_\pm $. Only the magnitudes $|\tilde \Delta_\pm|$ are important for the topological classification in the inverted regime ($\delta>0$). Clearly, this conclusion holds equally for the case $m=3$ and $C_6$ symmetry; in particular, Eq.~\eqref{eq:Delta+-} is still valid.

\subsection{Spin angular momentum models \label{ssec:spinful}}

The two-band models constructed in the previous subsection all rely on on-site orbital states with integral angular momentum. This property is not strictly required by Eq. \eqref{eq:Lz}, since it only fixes the relative angular momentum. Therefore, a different approach to engineering a band inversion of states with relative angular momentum $l$ relies on exploiting the spin degree of freedom. For instance, two states with spin quantum number $j_z = \pm l /2$ with odd $l$ have relative angular momentum $l$. Similarly, by considering states with general spin quantum numbers $l_1/2$ and $l_2/2$, and engineering couplings between such states, it becomes possible to realize band inversions with angular momentum $(l_1-l_2)/2$, where $l_{1,2}$ are both odd. In this subsection we follows this approach and present a number of simple spinful models which realize band inversion transitions to Chern insulators with higher Chern number.

In the presence of a spin degree of freedom a minimal model describing a band inversion must have four bands. We therefore begin by considering a triangular lattice model with two spin $j_z = \pm \frac32$ Kramers pairs. We introduce the electron operators $c_{\bk \Up,\Down}$ for each Kramers pair, where $\Up,\Down \equiv \pm \frac32$, and collect these in a vector $c_\bk$ defined as
\be
c_\bk = \begin{pmatrix} c_{\bk \Up \alpha} \\  c_{\bk \Down \alpha} \end{pmatrix}.   \label{eq:spin-3/2}
\ee
Here $\alpha=1,2$ is a flavor index which labels the two pairs. The Hamiltonian is then defined as $H = \sum_\bk c^\dagger _\bk h_\bk c_\bk$ with four-band Hamiltonian matrix $h_\bk$. To describe the couplings between spin states we introduce a set of spin Pauli matrices $\sigma_{x,y,z}$, where $\sigma_z=\pm 1$ corresponds to $\Up,\Down$; we use the Pauli matrices $\tau_{x,y,z}$ to describe couplings in flavor space. 

The form of the Hamiltonian $h_\bk$ can be determined using the same symmetry prescription as before. The symmetry of the spin matrices $\sigma_{x,y,z}$ follows from the transformation properties of the $j_z = \pm \frac32$ spin states, which are different from the transformation properties of a more familiar $j_z = \pm \frac12$ doublet. In particular,
the spin matrices $\sigma_x$ and $\sigma_y$ do not transform as the $x,y$-components of an $S=1$ angular momentum, which transform as $p_{x,y}$ waves, but instead transform as $f$ waves. This implies that a rotationally symmetric coupling of the spin states has $f$-wave symmetry. We again take $\varepsilon_\bk = 3 - \lambda^s_\bk - \delta$ and find that a minimal Hamiltonian with $C_6$ symmetry but broken $T$ and $M$ symmetry takes the form
\be
 h_\bk  = \varepsilon_\bk \tau_z +b_z \sigma_z+ \Delta_1\lambda^{f_1}_\bk\tau_x\sigma_x +\Delta_2\lambda^{f_2}_\bk\tau_x\sigma_y.   \label{eq:H-spin-triangle-f}
\ee
%
The first term describes the dispersion and energy difference $\delta$ of two spin-degenerate bands. Here, we are interested the regime where these bands remain uninverted and therefore set $\delta<0$. The second term describes a Zeeman splitting of the $j_z = \pm \frac32$ Kramers pair states in each band, and as such it breaks $T$, vertical reflections, and twofold rotations about in-plane axes; the Zeeman splitting preserves $C_6$. For Eq.~\eqref{eq:H-spin-triangle-f} to describe a band inversion with angular momentum $l=\pm 3$, we consider the case $|b_z|>\delta$, which corresponds to an inversion of a $j_z=\frac32$ and $j_z=-\frac32$ band with different flavor index. The final two terms then describe an $f$-wave coupling between the spin species which is off-diagonal in flavor space, i.e., connects the different Kramers pairs. This coupling gaps out the inverted bands and realizes a Chern insulating phase in the way described by Eq.~\eqref{eq:H}. We note here that the $f$-wave coupling of Eq.~\eqref{eq:H-spin-triangle-f} does not break $T$ and can thus viewed as a form of spin-orbit coupling; we will return to this observation in Sec.~\ref{sec:T-invariant}. This remains true when considering a slightly more general coupling of the form $  (\Delta_1\lambda^{f_1}_\bk +\Delta_2\lambda^{f_2}_\bk)\tau_x\sigma_++\text{h.c.} $, where $\Delta_{1,2}$ are complex. The latter form should be viewed in the context of the discussion of Eqs.~\eqref{eq:Delta-k-C4} and \eqref{eq:Delta+-}.

Next, consider the case of two Kramers pairs with $j_z = \pm \frac32$ and $j_z = \pm \frac12$, respectively. Adopting the notation $\up,\down \equiv \pm \frac12$ for the two $j_z = \pm \frac12$ states, we can collect the electron operators in a vector $c_\bk$ given by
\be
c_\bk = ( c_{\bk \Up}, c_{\bk \up}, c_{\bk \down} ,  c_{\bk \Down} )^T ,     \label{eq:j=3/2}
\ee
which has the structure of a $j=\frac32$ quartet. Since the particle-hole pairs $c^\dagger_{\bk \Up} c_{\bk \down} $ and $c^\dagger_{\bk \Down} c_{\bk \up} $ have angular momentum $+2$ and $-2$, respectively, we can seek to engineer a band inversion between the corresponding bands and couple these with angular momentum $l=\pm2$ pairing terms. The minimal Hamiltonian which achieves this has a structure similar to Eq. \eqref{eq:H-spin-triangle-f} and takes the form
\be
h_\bk =  \varepsilon_\bk \sigma_z\tau_z  +b_z \sigma_z + \Delta(\lambda^{d_+}_\bk\sigma_- +\lambda^{d_-}_\bk\sigma_+).    \label{eq:H-spin-triangle-d}
\ee  
Here  $\tau_{z} = \pm1$ still describes the two Kramers pairs but the basis is defined by Eq. \eqref{eq:j=3/2}. As in Eq. \eqref{eq:H-spin-triangle-f}, the first two terms are responsible for the band inversion and the final two terms describe a $d$ wave pairing of the inverted bands, giving rise to an energy gap. Recall that the $d$ waves are degenerate on the triangular lattice, leading to a single $\Delta$. Due to the $d$-wave nature of the coupling the ground state of \eqref{eq:H-spin-triangle-d} realizes a Chern insulator with $C=\pm 2$.

\section{Interactions and excitonic pairing \label{sec:interactions}}

In this section we turn to a more thorough study of the Chern insulator models introduced in the previous section. In particular, we address the effect of electronic correlations on the nature of the band inversion transition. As explained in Sec.~\ref{sec:pairing}, the Hamiltonian of Eq.~\eqref{eq:H} describes a band inversion transition of non-interacting fermions. Similarly, the lattice models introduced in the previous section are free fermion models. To see how interactions can affect the nature of the band inversion, consider the critical point defined by $\delta=0$ where the two bands touch at $\bk=0$. First note that symmetry protects the quadratic dispersion of the bands at the touching point, which implies that the density of states does not vanish. This should be contrasted with a Dirac fermion transition, characterized by linear dispersion at the touching point, for which the density of states vanishes. Due to the nonzero density of states it is natural to expect that interactions give rise to correlated states with an energy gap. 

Two different possibilities for correlated states can be distinguished. The first is the formation of an excitonic insulator defined by the condensation of (conduction band) electron and (valence band) hole bound states. The condensation of electron-hole excitons breaks rotational symmetry and is therefore associated with a spontaneously broken (discrete) symmetry. The second possibility is the formation of a correlated liquid of electrons and holes which does not break symmetries but instead has fractional quantum Hall topological order~\cite{dubail2015,hu2018}. This intriguing second scenario has motivated a previous study~\cite{hu2018}, in which we proposed and analyzed a wave function description for such correlated liquid of electrons and holes. In this work we focus on the first scenario and study the excitonic insulator state in the vicinity of the band inversion. More precisely, we consider the mean field theory of the excitonic insulator.

We have argued in Sec.~\ref{sec:pairing} that the description of the higher angular momentum band inversions is formally similar to the BCS theory for (higher angular momentum) pairing states of fermions. In case of the former, however, there is no notion of a broken symmetry in the absence of interactions. The interaction-driven excitonic insulator, on the other hand, does break a symmetry and its mean field theory (at low-energies) is an analog of BCS theory for $s$-wave pairing. As a result, the formation of excitons can be referred to as excitonic pairing of electrons and holes. As we will demonstrate, the development of a mean field theory for excitonic pairing, in close analogy with BCS theory, gives access to information on the structure of the ground state in the vicinity of the band inversion transition. Most importantly, this will lead us to the conclusion that the ground state in the band inverted regime can be viewed as a multicomponent $C=1$ quantum Hall liquid of electrons and holes.

It is worth pointing out that the present case of quadratically crossing bands is different from previously studied quadratic band crossing models~\cite{sun2009,uebelacker2011,murray2014,dora2014}. In the latter the degeneracy at the touching point is protected by point group and $T$ symmetry. In contrast, in the present case the touching point is not symmetry-protected, but instead defines the critical point of the band inversion transition parametrized by $\delta$; $\delta$ does not reflect a broken symmetry.

\subsection{Excitonic insulator mean field theory \label{ssec:meanfield}}

\subsubsection{General analysis of the continuum model  \label{sssec:mf-continuum}}

\begin{figure}
\includegraphics[width=0.9\columnwidth]{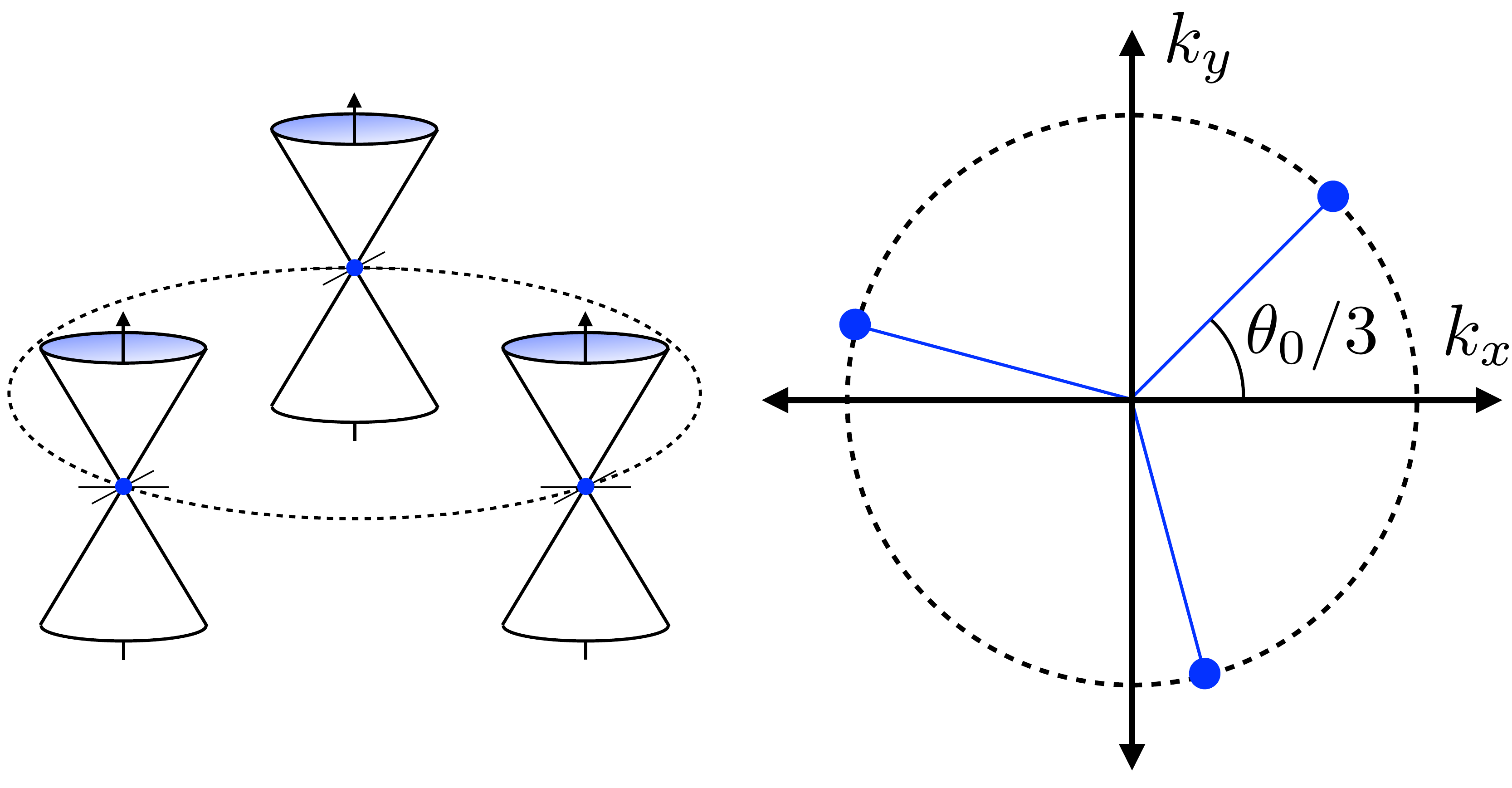}
\caption{\label{fig:dirac-points} {\bf Dirac points at the topological phase transition.} For the case $m=3$ the transition from the topological Chern insulating phase to the excitonic insulator phase is marked by three Dirac points, as shown schematically on the left. This transition is described by Eq.~\ref{eq:Delta-C2-C3} and occurs when $|\Delta_0| =|\Delta_{m=3}|(2\delta)^{3/2}$. The Dirac points are located on a circle with radius $k_F$ and are related by threefold rotation symmetry, as shown on the right. Importantly, the angle at which the Dirac points are located is determined by the phase of  $\Delta_0$. 
 }
\end{figure}

To begin, consider the low-energy description of the square and triangular lattice models of Eqs. \eqref{eq:hkC4} and \eqref{eq:hkC6}. In the analysis that follows we particularize to these models for illustrative purposes, without loss of generality. Consider furthermore the special case $\Delta_1=\Delta_2\equiv \Delta$; according to Eq. \eqref{eq:Delta-expand}, for small momenta $\bk$ this implies $\Delta_\bk  \sim \Delta_m (k_x-ik_y)^m$ with $m=2,3$. Based on Eq.~\eqref{eq:Delta+-}, we promote $\Delta_m$ to a complex number with arbitrary phase. In addition, in the small momentum limit one has $\varepsilon_\bk \simeq \bk^2/2-\delta$. As discussed, the form of $\Delta_\bk$ (i.e., an eigenstate of $L_z$ with angular momentum $l=-m$) is determined by the rotational symmetry of the system. Importantly, the formation of excitons, i.e., excitonic pairing, alters the form of $\Delta_\bk$ and breaks rotational symmetry. Specifically, in a mean field description of excitonic pairing $\Delta_\bk$ becomes 
\be
\Delta_\bk = \Delta_0 + \Delta_m(k_x -ik_y)^m,  \label{eq:Delta-C2-C3}
\ee
where $m=2,3$ and $\Delta_0$ represents the formation of excitons. We observe that $\Delta_0$ is an angular momentum $l=0$ coupling of conduction and valence band, and since \eqref{eq:Delta-C2-C3} is a superposition of terms with different angular momentum, rotational symmetry is broken. In the low-energy continuum limit the $l=0$ angular momentum term breaks the emergent continuous rotation symmetry and lowers the symmetry to $\mathbb{Z}_m$. In particular, $\Delta_0$ transforms as $\Delta_0 \to e^{-im\theta}\Delta_0$ under rotations by an angle $\theta$. This establishes a link between the phase of $\Delta_0$ and rotation symmetry breaking, which is analogous to the link between the superconducting phase and $U(1)$ charge conjugation. 

On the lattice, in the case $m=2$ the fourfold rotation $C_4$ is reduced to $C_2$; in the case $m=3$ the rotational symmetry is lowered from $C_6$ to $C_3$. In both cases, $m=2$ and $m=3$, the form of \eqref{eq:Delta-C2-C3} can be derived from a lattice model mean field Hamiltonian given by $h_\bk \to h_\bk + \Delta_0 \tau_++ \Delta^*_0 \tau_-$. 

To examine the implications of \eqref{eq:Delta-C2-C3}, in particular the excitonic term, it is useful to invoke the connection to the problem of pairing states. In the context of pairing states, $\Delta_0$ can be interpreted as an $s$-wave pairing. Assuming one is in the band inverted regime, this implies a transition as function of the strength of $\Delta_0$, from a Chern insulating phase to a trivial insulator phase. This implication follows from the fact that $s$-wave pairing is topologically trivial. The transition occurs when $|\Delta_0| = |\Delta_m| k^m_F =  |\Delta_m| (2\delta)^{m/2}$, where $k_F$ is a momentum defined by the condition $\varepsilon_\bk=0$ (see Sec.~\ref{sec:pairing}). At the transition the system is gapless, with three ($m=3$) or two ($m=2$) Dirac points located in a circle in momentum space with radius $k_F$. Thus, the transition is marked by three (or two, in the case of $m=2$) Dirac fermion mass inversions, which is consistent with the total change in the Chern number. This is shown schematically in Fig.~\ref{fig:dirac-points} for the case $m=3$. Note that the location of the Dirac points depends on the phase of $\Delta_0$: assuming $\Delta_0=|\Delta_0|e^{i\theta_0}$ and $\Delta_m$ real but negative, the Dirac points are located at angles $\theta_0/m +j2\pi/m$ with $j=0,1,2$.

\subsubsection{Mean field phase diagram  \label{sssec:mf-lattice}}

Having discussed the qualitative features of the excitonic mean field theory, we now turn to a more quantitative analysis. To this end, we take the triangular lattice model of Eq.~\eqref{eq:hkC6} (the analysis is similar for the $m=2$ square lattice model), in which we set $\Delta_1=\Delta_2\equiv\Delta_{m=3}$, and add an onsite Hubbard repulsion of the form $H_U = U \sum_j n_{js}n_{jf}$, where $n_{s,f}$ are the density operators of the $s$ and $f$ orbitals and the sum is over sites. In momentum space the Hubbard repulsion takes the form
\be
H_U = \frac{U}{N}\sum_{\bq} \sum_{\bk\bk'} s^\dagger_{\bk+\bq} s_\bk f^\dagger_{\bk'} f_{\bk'+\bq},  \label{eq:H-U-k}
\ee
where $N$ is the system size (i.e., total number of sites). By performing a mean field decoupling of \eqref{eq:H-U-k} in the excitonic channel (see Appendix \ref{app:mftheory} for details) one obtains a self-consistency condition for the excitonic order parameter $\Delta_0$ given by
\be
\Delta_0 = -\frac{U}{2N}\sum_{\bk} \langle   \varphi^\dagger_\bk \tau_x\varphi_\bk \rangle.  \label{eq:mean-field}
\ee
Here $\varphi_\bk$ are the fermion operators defined in Eq.~\eqref{eq:basisC6}. At zero temperature Eq.~\eqref{eq:mean-field} defines the stationary point of the free energy density 
\be
F[\Delta_0]  = - \sum_{\bk} E_\bk + \frac{N}{U}\Delta^2_0 ,  \quad  E_\bk = \sqrt{\varepsilon^2_\bk + |\Sigma_\bk|^2}, \label{eq:F}
\ee
where $\Sigma_\bk$  is defined as $\Sigma_\bk = \Delta_0 +\Delta_\bk$ with $\Delta_\bk = \Delta_{3}(i \lambda^{f_1}_\bk+\lambda^{f_2}_\bk) $. 


\begin{figure}
\includegraphics[width=\columnwidth]{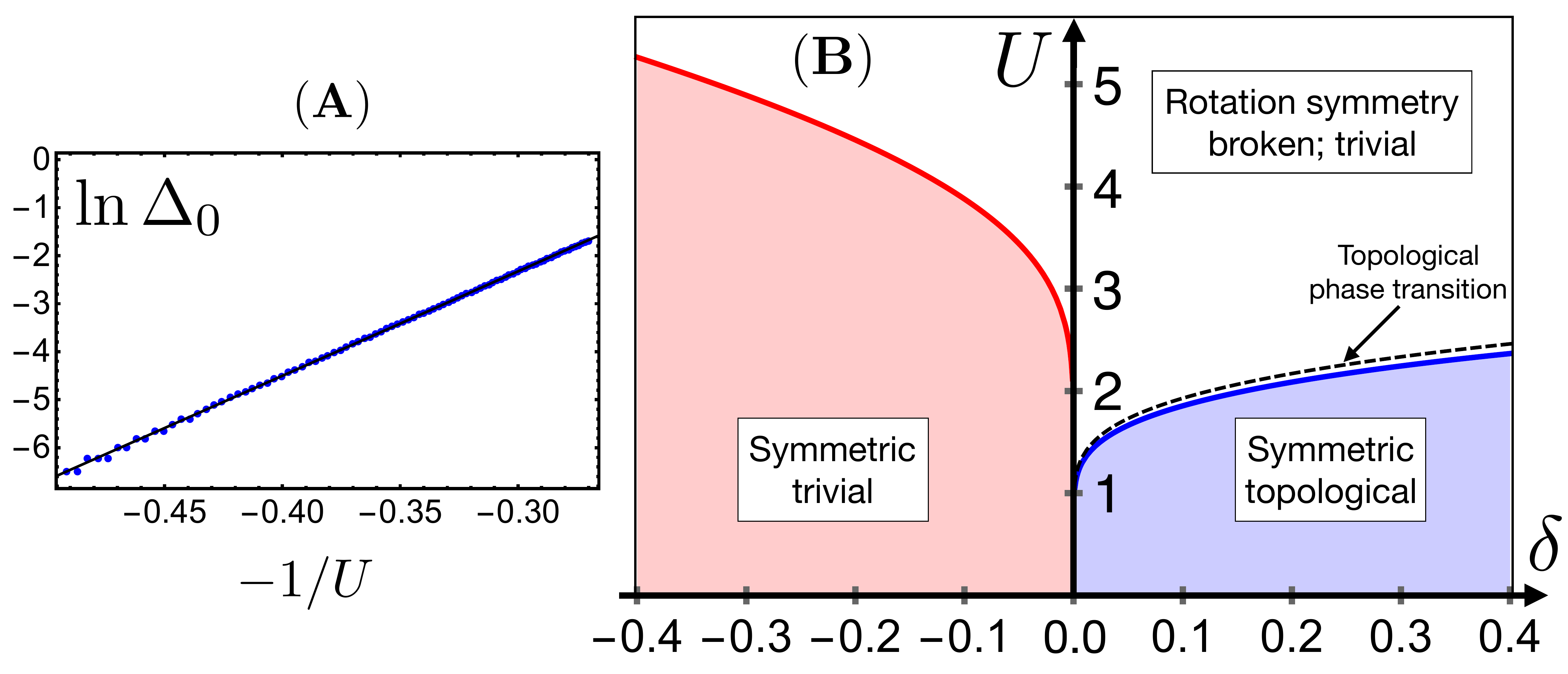}
\caption{\label{fig:MF-theory} {\bf Excitonic mean-field theory.} Panel ($\bf A$) shows the dependence of $\Delta_0$ on $U$ at the band inversion transition defined by $\delta=0$; since $\Delta_0 \sim \exp({-1/\alpha U})$, we plot $\ln \Delta_0$ as function of $-1/U$. Panel ($\bf B$) shows the $U$--$\delta$ phase diagram obtained from the mean field theory of excitonic pairing at zero temperature. In the inverted regime ($\delta>0$) the blue curve shows the phase boundary separating the rotation symmetric topological phase from the rotation symmetry broken phase with nonzero excitonic pairing. The critical interaction strength $U_c$, which defines this phase boundary, is obtained from \eqref{eq:U_c}. The dashed curve indicates the presence of a second transition in the vicinity of the symmetry breaking transition, at $U'_c>U_c$, which separates the symmetry broken topological phase from the trivial excitonic insulator. In all calculations the overall energy scale is fixed by setting $t=1$ and $\Delta_3$ is set to $\Delta_3=1.0$. 
 }
\end{figure}

Solving these equations at zero temperature, we obtain a phase diagram of excitonic pairing as function of the interaction strength $U$ and the band inversion parameter $\delta$. The results are presented in Fig.~\ref{fig:MF-theory}, which we now discuss. We first focus on the case $\delta=0$. In this case, the non-interacting systems is right at the topological transition and is gapless, with two quadratically dispersing bands touching at $\bk=0$. As a consequence of the non-vanishing density of states at the gapless point, the susceptibility is divergent and one expects a rotation symmetry broken state with nonzero $\Delta_0$ at infinitesimal $U$. More precisely, one expects $\Delta_0 \sim \exp(-1/\alpha U)$, where $\alpha$ is a constant reflecting the density of states \cite{sun2009}. This is confirmed in Fig.~\ref{fig:MF-theory} ($\bf A$), where we show $\ln \Delta_0$ as function of $-1/U$ for the case $\delta=0$. (In all calculations we choose $\Delta_3=0.5t$.) 

We then proceed to the case $\delta \neq 0$. For nonzero $\delta$, when the non-interacting system given by $h_\bk$ in Eq. \eqref{eq:hkC6} is gapped, one expects a transition to the rotation symmetry broken state at finite interaction strength $U_c$. The critical interaction strength as function of $\delta$ defines the phase boundary which separates the rotation symmetric phase from the rotation symmetry broken phase with nonzero excitonic pairing. Since the inverted regime ($\delta>0$) and the uninverted regime ($\delta<0$) have different dispersion, as is clear from Fig.~\ref{fig:band-inversion} ($\bf A$) and ($\bf C$), the critical strength $U_c$ is expected to be smaller in the inverted regime. We find that the transition to the symmetry broken phase is a second order transition in mean field theory, which implies that a closed form expression for $U_c$ can be obtained by expanding $F$ of Eq.~\eqref{eq:F} in powers of $\Delta_0$. Such Landau-type expansion can only have even powers of $\Delta_0$ and is valid in the vicinity of the transition when $\Delta_0$ is small; specifically, one has up to fourth order
\be
F[\Delta_0]/N  = \left(U^{-1}-c_2\right) \Delta^2_0 +c_4 \Delta^4_0.  \label{eq:F-Landau}
\ee
Then, $U_c$ is defined by the condition that the coefficient of the quadratic term vanishes and we find
\be
U_c = \frac{1}{c_2}, \quad c_2 = \frac{1}{2N}\sum_\bk \left[\frac{1}{E_\bk} - \frac{(\text{Re}\Delta_\bk )^2}{E^3_\bk} \right],  \label{eq:U_c}
\ee
where $E_\bk$ defined in Eq.~\eqref{eq:F} is evaluated at $\Delta_0=0$. Figure~\ref{fig:MF-theory} ($\bf B$) shows the $U$--$\delta$ phase diagram obtained by evaluating $U_c$ as function of $\delta$. As expected, $U_c$ is smaller in the inverted regime (blue curve) compared to the uninverted regime (red curve). 

As discussed above in Sec.~\ref{sssec:mf-continuum}, in the inverted regime defined by $\delta>0$ one expects a second transition as the interaction strength increases. This second transition is a topological phase transition described by three Dirac fermions and occurs when $|\Delta_0| \sim |\Delta_{\bk = k_F}|$. One may thus identify a second $U'_c$ associated with the topological phase transition and it is then natural to ask how $U'_c$ differs from $U_c$. To get an understanding we employ the Landau theory of Eq.~\eqref{eq:F-Landau} and solve for $\Delta_0$. Minimization directly yields $|\Delta_0| = \sqrt{(c_2-U^{-1})/2c_4}$. Within this approach the value of $U'_c$ is determined by setting this result equal to the value of $\Delta_0$ at which which the topological transition occurs. Defining the latter as $\tilde \Delta_0$, we find 
\be
\frac{U'_c-U_c}{U_c} = \frac{1}{(2c_4U_c \tilde \Delta^2_0)^{-1} -1 }.  \label{eq:U'_c}
\ee
We have verified that this estimate based on \eqref{eq:F-Landau} is in good agreement with the numerically exact result. Since $(2c_4U_c \tilde \Delta^2_0)^{-1} \gg 1 $, Eq.~\eqref{eq:U'_c} implies that the transitions, i.e., the symmetry breaking transition and the topological transition, are in close proximity. This is indicated by the dashed line in Fig.~\ref{fig:MF-theory} ($\bf B$). To understand why $(U'_c-U_c)/U_c$ is small, it is helpful to consider the continuum description discussed in Sec.~\ref{sssec:mf-continuum}, which is valid for small $\delta$. In this case one has $\tilde \Delta^2_0 \sim \delta^3$ and $c_4 \sim \delta^{-2} $, from which one finds $(U'_c-U_c)/U_c \sim \delta / \ln \delta $. Note that in case of the square lattice one finds $(U'_c-U_c)/U_c \sim 1 / \ln \delta $.

The close proximity of the two transitions is an interesting aspect of higher angular momentum band inversions. The fate of these two transitions in an interacting theory beyond mean field will be an interesting question to address. Such theory should be formulated in terms of three flavors of Dirac fermions coupled to a fluctuating phase of the excitonic order parameter $\Delta_0$, as suggested by Fig.~\ref{fig:dirac-points}.

\subsection{Structure of the ground state \label{ssec:GS}}

Having discussed the quantitative aspects of the excitonic pairing mean field theory, we return to a more conceptual analysis, which we develop within the low-energy continuum model. More specifically, we proceed to examine the structure of the ground state as defined in Eq.~\eqref{eq:Phi-GS}. As discussed in Sec.~\ref{sec:pairing} (see also Appendix~\ref{app:pairing}), the continuum model ground state is specified in terms of the function $g_\bk$ and we demonstrate below that the excitonic pairing term $\Delta_0$ plays a key role in the interpretation of $ g(\br)$, the Fourier transform of $g_\bk$. This leads to the conclusion that a theory for the band inversion transition which includes the rotation symmetry breaking excitonic pairing term gives access to the structure of the electron-hole ground state. 

To show this, it is useful to first consider the case $\Delta_0=0$, i.e., when rotation symmetry is not broken, and obtain $ g(\br)$. We note that the form of $g_\bk$, and thus $g(\br)$, changes across the band inversion transition and thus depends on $\delta$. We focus on two cases: the critical point of the transition when $\delta=0$, and the Chern insulating phase when $\delta>0$. Consider the former case first. Right at the transition and for small momentum $\bk\to0$ one finds that $g_\bk \propto \bk^2 /\Delta(k_x -ik_y)^m$. Taking the Fourier transform to obtain $ g(\br)$ one obtains 
\be
 g(\br) \propto      1/ z^m  ,  \label{eq:g(r)-1}  
\ee
where $z=x+iy$. Note that since $g_\bk$ is considered in the small momentum limit, \eqref{eq:g(r)-1} describes the long-distance behavior of $ g(\br)$. In this limit $ g(\br)$ falls off as a power law as function of the distance between the electron and hole forming a pair, and this regard the interpretation of \eqref{eq:g(r)-1} as describing the pairing of electrons and holes with angular momentum $l=-m$ makes sense. Furthermore, in Ref.~\onlinecite{hu2018} wave argued that many-body Slater-determinant of \eqref{eq:g(r)-1}, defined by \eqref{eq:Phi-GS}, can be related to lowest Landau level wavefunctions at filling factor $\nu=1/m$. This argument was based on a comparison of the number of zeros of the many-body wave function, viewed as a function of one of its variables. 


Consider next the band-inverted regime $\delta>0$ (still taking $\Delta_0=0$). In this case the small momentum limit of $g_\bk$ is given by $g_\bk \propto \delta/\Delta^*(k_x +ik_y)^m$, which implies that in the long-distance limit $ g(\br)$ has the form
\be
 g(\br) \propto (z^*)^{m-1}/z.     \label{eq:g(r)-2} 
\ee
As a result, one has that $|g|$ is constant for $m=2$~\cite{read2000} and $|g| \sim |z|$ for $m=3$ at long distances. This long-distance behavior of $ g(\br)$ presents a puzzle, since it is not immediately clear how to reconcile it with the interpretation of $ g(\br)$ as describing the pairing of electrons and holes; the electrons and holes cannot be said to be bound into a pair in a meaningful sense. In contrast, this is different for the well-known case of an $l=-1$ band inversion transition described by a Dirac fermion, which corresponds to $m=1$ in \eqref{eq:g(r)-2}. In the case of the latter, \eqref{eq:g(r)-2} corresponds to the ``weak-pairing phase''~\cite{read2000} and the many-body Slater determinant of the electron-hole paired state defines a many-body wavefunction for the $C=\pm1$ Chern insulator. 

To make progress in understanding the higher Chern number phases in the inverted-regime, we break rotation symmetry by introducing a nonzero $\Delta_0$, such that $\Delta_k$ is given by \eqref{eq:Delta-C2-C3}. As explained earlier, the amplitude of $\Delta_0$ controls a transition from a Chern insulating phase to a trivial insulator phase, while keeping $\delta$ fixed. As far as the topology of the two phases is concerned, this is same topological transition as the transition controlled by $\delta$ (while keeping $\Delta_0=0$). The former, however, is characterized by a critical gapless phase with three linearly dispersing Dirac points and as a result, the topological transition parametrized by $\Delta_0$ is described by $m$ simultaneous $l=-1$ band inversions, consistent with a total angular momentum $l=-m$ transition. Each of these $l=-1$ band inversions, which are described by a Dirac fermion theory, is well understood and has $g(\br) \propto 1/z$. Consequently, the band inversion via three Dirac points reveals that a higher angular momentum band inversion has the generic structure of three $l=-1$ Dirac fermion transitions, with three flavors of electron-hole pairing states describing a $\nu=1$ quantum Hall phase. 

This argument can be put on a more precise footing by considering $g_\bk$ for $\Delta_k$ given by \eqref{eq:Delta-C2-C3}. In this case one finds
\be
g_\bk \propto \frac{\delta}{k^m_+ + \Delta_0/\Delta_m},     \label{eq:gk-Delta0} 
\ee
which has $m$ first order poles at $k_n$ ($n=1,\ldots,m$) rather than one $m$-th order pole at $\bk=0$. Defining $k_1=(\Delta_0/\Delta_m)^{1/m}$ one has $k_n = e^{i2\pi(n-1)/3}k_1$ and \eqref{eq:gk-Delta0} can be written as a sum over the three poles $ \sum_{n=1}^{m} \gamma_n/(k_+-k_n)$, where $\gamma_n$ are the residues. Fourier transforming then gives the expected form of $ g(\br)$ for three $l=-1$ transitions, with additional oscillatory factors originating from the nonzero momenta $k_n$.

A few comments are in order regarding the significance of rotational symmetry breaking. As explained, the existence of $m$ Dirac points at three distinct nonzero momenta requires the breaking of rotation symmetry. When full rotation symmetry is present it forces the three transitions to all occur at $\bk=0$, which in a sense obscures the topological structure of the transition, as evidenced by \eqref{eq:g(r)-2}. As far as the topological structure of the transition between the higher Chern number insulating phase and the trivial insulator is concerned, the presence of higher rotational symmetry is not required. In fact, from the perspective of topology the situation where the three $l=-1$ transitions occur at different momenta is more generic. 

A similar reasoning relying on broken rotation symmetry has been presented by Read and Green in the context of chiral $d$-wave pairing~\cite{read2000}, which may be compared to our $m=2$ case. In the case of chiral $d$-wave pairing, Read and Green showed that by studying the transition to a trivial $s$-wave pairing state---in contrast to changing the chemical potential---the correct edge excitation spectrum and vortex states of a chiral $d$-wave superconductor can be obtained. Since both the edge and vortex modes are rooted in the topological structure of the phase, this is another instance where only the more generic transition described by multiple Dirac fermions (and with broken rotation symmetry) reveals the true nature of the phase. 


\subsection{$m$-component $C=1$ quantum Hall states \label{ssec:m-component}}

The previous analysis of band inversions with broken rotation symmetry, in particular the splitting into multiple $l=-1$ band inversions, leads to an important insight regarding the structure of higher Chern number phase. It can be states as follows: Since the transition is described by $m$ flavors of Dirac fermions, the higher Chern number phase can be viewed as an $m$-component $C=1$ phase, of which each component is characterized by a quantum Hall wavefunction for electron-hole pairs at the Dirac point. 

It should be emphasized that here we reach this conclusion by considering a theory for the band inversion transition and do not make reference to the notion of a full Chern band. This approach is very different from---but may be compared to---an approach which explicitly addresses the structure of the Chern band by studying its Wannier state representation~\cite{barkeshli2012}. The latter approach clearly requires knowledge of the full Chern band, as the Wannier state representation is inaccessible within a (low-energy) continuum model which addresses the band inversion. Using the Wannier state representation, Ref.~\onlinecite{barkeshli2012} showed that a band with Chern number $C>1$ can be mapped to $C$ layers of Landau levels, each of which is equivalent to a $C=1$ band. Even though the two approaches are different, we thus see that both point to a characteristic structural property of higher Chern number bands: they are intrinsically multi-component in nature, with the number of components given by the Chern number $C$.

The Wannier state representation of bands with higher Chern number leads a further important observation regarding the action of translational symmetry on the multi-layer quantum Hall systems under the mapping~\cite{barkeshli2012}. Due to the structure of the Wannier states, one of the two primitive translations acts as a permutation on the $C$ layers and thus acts nontrivially on the layer degree of freedom. This was shown to have rather drastic consequences when lattice dislocations are present. In particular, dislocations give rise to an intricate interplay between geometry and topology, resulting in topological degeneracy even for Abelian states. 

In the framework of the continuum model for the band inversion transition, we can establish a connection to this result by considering the effect of the $m$-fold rotations. As noted earlier, the $m$-fold rotations give rise to a residual $\mathbb{Z}_m$ symmetry. Furthermore, the $m$-fold rotations permute the $m$ Dirac points and thus permute the $m$ $C=1$ components. As an example, consider $m=3$ and let $\bk_{0,1,2}$ denote the location of the three Dirac points at the transition, as shown in Fig.~\ref{fig:dirac-points}. The threefold rotation relates these as $\bk_{n} = C^n_3\bk_0$, where $n=0,1,2$. As explained in Sec.~\ref{sssec:mf-continuum}, the three Dirac point momenta $\bk_{0,1,2}$ are determined by the phase of $\Delta_0$. A $U(1)$ vortex in the phase of $\Delta_0$ is associated with a $2\pi/3$ rotation and permutes the Dirac points. This suggests an interesting field theoretic description of the band inversion transition in terms of an $XY$ variable $\Delta_0$ and three Dirac fermions, where proliferation of vortices in the phase of $\Delta_0$ restores rotational symmetry and leaves the Dirac fermions ill-defined. We leave the systematic development and analysis of such field theoretic description of higher angular momentum band inversions for future studies.

\section{Time-reversal invariant generalizations  \label{sec:T-invariant}}

\subsection{Transition from normal to topological insulator  \label{ssec:NI-TI}}

Now that we have introduced a class of Chern insulator models based on the notion of higher angular momentum band inversions, both the theory and the historical development of topological insulators lead to a natural question: do there exist time-reversal invariant generalizations of such models? For the orbital models of Sec.~\ref{ssec:spinless} the answer is clearly yes, since we may simply introduce a spin degree of freedom and build a $T$-invariant Hamiltonian by combining two copies of $h_\bk$: one for the up spins and a time-reversed version of $h_\bk$ for the down spins. In particular, in the spirit of BHZ \cite{bernevig2006} one can define 
\be
\mathcal H_\bk  = \begin{pmatrix} h_\bk & \\  & h^*_{-\bk} \end{pmatrix}.  \label{eq:BHZ}
\ee
This Hamiltonian describes a transition between a trivial insulator and a Chern insulating phase in each spin sector, where the Chern numbers associated with the spin species have opposite sign. This can be viewed as a transition between a normal insulator and a topological insulator characterized by an integer number of helical edge modes. The number of helical edge modes is equal to the angular momentum of the transition.

At low-energies, close to the band inversion transition, the coupling of the $\ket{l=0,\pm \frac12}$ and $\ket{l=\pm m,\pm \frac12}$ bands is a diagonal matrix $\Delta_\bk$ in spin space given by
\be
\Delta_\bk  = \Delta \begin{pmatrix} k_\pm^m & \\  & (-k_\mp)^m \end{pmatrix},  \label{eq:Delta-k-spin}
\ee
with $m=2,3$ and $k_\pm = k_x\pm ik_y$. By construction, this implies that the transition from normal to topological insulator (or vice versa) is special in the sense that right at the critical point of the transition (i.e., when the bands touch) the bands disperse quadratically. As in Sec.~\ref{sec:interactions}, one then expects interaction effects to be important. In this time-reversal invariant case, the two possibilities for correlated states are the excitonic insulator and the fractional topological insulator~\cite{bernevig2006b,levin2009,karch2010,lu2012,chan2013,maciejko2015,neupert2015,neupert2011b,repellin2014}. In particular the fractional topological insulator is an interesting possibility, and band inversions of the type described by \eqref{eq:BHZ} and \eqref{eq:Delta-k-spin} are a promising venue for their realization. 

The Hamiltonian of Eq.~\eqref{eq:BHZ} has the property that it commutes with spin rotations about the $z$-axis, i.e., $[\mathcal H_\bk, \sigma_z ]=0$, which implies that $S_z$ is conserved. This property, however, is not guaranteed unless it is mandated by appropriate physical symmetries of the system. For a given symmetry group the most general Hamiltonian allowed by symmetry may have spin-orbit coupling terms which violate $S_z$ conservation. Since such terms are likely to spoil the form of the coupling $\Delta_\bk$ at low energies, and thus potentially destroy the preconditions for interactions to be important, it is necessary to determine under what conditions the form of \eqref{eq:BHZ} is enforced by symmetry.

\subsection{Symmetry protection  \label{ssec:NI-TI}}

To examine the symmetry protection of the $T$-invariant band inversion, we consider the axial point groups of two-dimensional layer groups (as in Sec.~\ref{sec:models}) and determine the constraints each imposes. Importantly, whereas in Sec.~\ref{sec:models} we only needed to consider symmetry groups compatible with nonzero chirality, here we must consider a more general class of axial symmetry groups. These groups are summarized in Table~\ref{tab:sym-protect}, organized by crystal system and the presence of inversion symmetry. 

We start by examining systems with orbital ($l$) and spin ($j_z$) degrees of freedom given by $(l,j_z)=(m,\pm \frac12)$, with $m=2,3$, which are simple spinful generalizations of the models introduced in Sec.~\ref{ssec:spinless}. We then consider $T$-invariant generalizations of the models introduced in Sec.~\ref{ssec:spinful}, which are constructed from two $j_z=\pm \frac32$ Kramers pairs. We conclude by discussing a generalized Kane-Mele model~\cite{kane2005b} based on Sec.~\ref{sssec:honeycomb}.

\subsubsection{Systems with $(l,j_z)=(m,\pm \frac12)$ states}

Consider the triangular lattice case with $s$ and $f$ states (i.e., $m=3$). We introduce the spin degree of freedom by defining
\be
H=\sum_\bk  \Phi^\dagger_\bk \mathcal H_\bk\Phi_\bk ,  \quad \Phi_\bk = \begin{pmatrix}  s_{\bk \up,\down  } \\   f_{\bk \up,\down } \end{pmatrix},  \label{eq:basisC6-spin}
\ee
such that $\mathcal H_\bk$ is matrix in orbital and spin space; $\sigma_z=\pm 1$ denotes $\up,\down$. A $T$-invariant version of Eq. \eqref{eq:hkC6} is given by
\be
\mathcal H_\bk = \varepsilon_\bk \tau_z + \Delta_1\lambda^{f_1}_\bk\tau_y +\Delta_2\lambda^{f_2}_\bk\tau_x\sigma_z,   \label{eq:hk-D6h}
\ee
which is clearly of the form \eqref{eq:BHZ}. To determine what symmetries are sufficient to protect the structure of the Hamiltonian we begin by examining the hexagonal and trigonal symmetry groups of Table~\ref{tab:sym-protect} with inversion symmetry. In the presence of both $T$ and inversion symmetry all bands are necessarily twofold degenerate, imposing a strong constraint on the Hamiltonian. 

We first observe that \eqref{eq:hk-D6h} is invariant under all symmetries of the hexagonal group $D_{6h}$. In fact, if $D_{6h}$ is imposed \eqref{eq:hk-D6h} exhausts all symmetry-allowed terms, which implies that the full group $D_{6h}$ is sufficient to protect the band inversion transition. The same is true for the trigonal group $D_{3d}$, which is a subgroup of $D_{6h}$. We conclude that both $D_{6h}$ and $D_{3d}$ protect a $T$-invariant band inversion of spinful $s$ and $f$ bands.

Next, consider the symmetry groups $C_{6h}$ and $S_{6}$. These differ from the previous two groups by the absence of twofold rotations about axes in the plane. As a result of the lower symmetry, the Hamiltonian takes a more general form given by
\begin{multline}
\mathcal H_\bk  = \varepsilon_\bk \tau_z + (\Delta_1\lambda^{f_+}_\bk +\Delta^*_1\lambda^{f_-}_\bk )\tau_y \\ 
+(\Delta_2\lambda^{f_+}_\bk +\Delta^*_2\lambda^{f_-}_\bk )\tau_x\sigma_z,   \label{eq:hk-C6h}
\end{multline}
where now $\Delta_{1,2}$ are complex and we have defined $\lambda^{f_\pm }_\bk =\lambda^{f_1}_\bk\pm i \lambda^{f_2}_\bk $. Since \eqref{eq:hk-C6h} still commutes with $\sigma_z$, the Hamiltonian is of the form \eqref{eq:BHZ}. The effect the of the more general coupling can be understood by expanding around the band inversion transition at $\bk=0$. We find
\be
\Delta_\bk =(-i \Delta_1 \pm \Delta_2 )k_+^3 + (-i \Delta^*_1 \pm \Delta^*_2 )k_-^3,  \label{eq:C6h-phase}
\ee
which should be compared to the discussion in Sec.~\ref{sssec:low-energy}. We see that the additional couplings only have an effect on the phase and amplitude of cubic terms and therefore do not fundamentally alter the structure of the band inversion.  As a result, all symmetry groups which possess an inversion symmetry provide sufficient protection for a $T$-invariant band inversion with higher angular momentum.

\begin{table}[t]
\centering 
\begin{ruledtabular}
\begin{tabular}{lccc}
  &  Hexagonal  & Trigonal & Tetragonal  \\
\hline  \\ [-2.0ex]
Inversion & $D_{6h}$  &  $D_{3d}$  &   $D_{4h}$   \\  [2pt]
                & $C_{6h}$  & $S_{6}$ &  $C_{4h}$   \\ [2pt] 
No Inversion  & $D_{6}$, $C_{6v}$, $D_{3h}$ & $D_{3}$, $C_{3v}$  &  $D_{4}$, $C_{4v}$, $D_{2d}$ \\   [2pt]
                      & $C_{6}$, $C_{3h}$ &  $C_{3}$  & $C_{4}$, $S_{4}$ 
\end{tabular}
\end{ruledtabular}
\caption{{\bf Classification of axial point groups}. Table summarizing the basic symmetry properties of the axial point groups. The point groups with an inversion symmetry can protect the structure of the band inversion given by Eqs.~\eqref{eq:BHZ} and \eqref{eq:Delta-k-spin}. Point groups on the second row differ from the first row by the lack of a twofold rotation perpendicular to the principal rotation axis; point groups on the fourth row differ from the third row by the lack of a perpendicular twofold rotation or a vertical mirror plane. }
\label{tab:sym-protect} 
\end{table}

We then proceed to the point groups listed in Table \ref{tab:sym-protect} which do not have an inversion symmetry. Owing to the absence of inversion symmetry, additional spin-orbit coupling terms can be symmetry-allowed. For instance, in the case of $C_{6v}$ the following two spin-orbit coupling terms are generically present in the Hamiltonian:
\be
 \mathcal H'_\bk  = t_1(\lambda^{p_1}_\bk\sigma_y - \lambda^{p_2}_\bk\sigma_x)+ t_2\tau_y(\lambda^{d_1}_\bk\sigma_y - \lambda^{d_2}_\bk\sigma_x) .   \label{eq:hk-C6v}
\ee
These terms do not commute with $\sigma_z$ and furthermore, when expanded in small momenta $\bk$, the first term describes a linear splitting of the spin species. Such linear coupling changes the nature of the band inversion, as it causes the density of states to vanish at the transition. A similar result is obtained for the symmetry groups $D_6$ and $C_6$, which leads to the conclusion that systems governed by these groups cannot have symmetry-protected higher angular momentum band inversions. The point group $D_{3h}$ is similar to $D_6$ and $C_{6v}$ but differs in an essential way: instead of a twofold rotation about the principal axis it contains a horizontal reflection. Since under the latter reflection $(\sigma_x,\sigma_y,\sigma_z)\to (-\sigma_x,-\sigma_y,\sigma_z)$ the terms of Eq.~\eqref{eq:hk-C6v} are symmetry-forbidden. We therefore find that $D_{3h}$ imposes sufficient constraints for the protection of the band inversion. This is not true for the point group $C_{3h}$, as its admits the coupling $\Delta_0 \tau_x$, which changes the nature of the band inversion transition.

Finally, since the trigonal groups without inversion are all subgroups of symmetry groups for which protection is lost, these do not protect the $T$-invariant higher angular momentum band inversion.   


We conclude this part by noting that a similar analysis applies to the square lattice Hamiltonian of Eq. \eqref{eq:hkC4}. Its $T$-invariant generalization based on \eqref{eq:BHZ} is given by
\be
 \mathcal H_\bk  = \varepsilon_\bk \tau_z + \Delta_1\lambda^{d_1}_\bk\tau_y +\Delta_2\lambda^{d_2}_\bk\tau_x\sigma_z.   \label{eq:hkC4spin}
\ee
The form of this Hamiltonian is protected by tetragonal $D_{4h}$ symmetry. Lowering the symmetry to $C_{4h}$ allows for the additional couplings $\Delta'_2\lambda^{d_2}_\bk\tau_y +\Delta'_1\lambda^{d_1}_\bk\tau_x\sigma_z$, which a similar effect as described by Eq.~\eqref{eq:C6h-phase}. When inversion symmetry is lacking, which is the case for symmetry groups $D_{4}$ and $C_{4v}$, the additional spin-orbit coupling term $\lambda^{p_1}_\bk\sigma_y - \lambda^{p_2}_\bk\sigma_x$ is activated.

\subsubsection{Systems with $j_z=\pm \frac32$ doublets \label{sssec:spin-3/2}}

In Sec.~\ref{ssec:spinful} we introduced models for higher angular momentum inversions of $j_z=\pm \frac32$ states. To describe such band inversions, it is necessary to consider two $j_z=\pm \frac32$ Kramers pairs, see Eq.~\eqref{eq:spin-3/2}. As a result, a $T$-invariant generalization can be obtained by imposing $T$ symmetry on the Hamiltonian defined in Eq.~\eqref{eq:H-spin-triangle-f}, which yields 
\be
 h_\bk  = \varepsilon_\bk \tau_z +  \tau_x(\Delta \lambda^{f_+ }_\bk \sigma_+ +\Delta^* \lambda^{f_- }_\bk \sigma_-),   \label{eq:H-spin-3/2}
\ee
where $\Delta$ is complex and $\lambda^{f_\pm }_\bk =\lambda^{f_1}_\bk\pm i \lambda^{f_2}_\bk $ as before. Close to $\bk=0$ the coupling between the bands is a matrix in spin space and reads as 
\be
\Delta_\bk  =  \begin{pmatrix}  &  \Delta k_\pm^3  \\   \Delta^* k^3_\mp  & \end{pmatrix}. \label{eq:D-k-spin-3/2}
\ee

To determine the symmetry protection it is necessary to specify the symmetry quantum numbers more precisely. Here we first assume the presence of inversion symmetry and focus on the case where one of Kramers pairs is inversion even and one is odd. This implies that \eqref{eq:H-spin-3/2} is invariant under inversion. More specifically, \eqref{eq:H-spin-3/2} is invariant under all symmetries of $D_{6h}$ and represents the most general form of the Hamiltonian with this symmetry. Furthermore, lowering the symmetry to $D_{3d}$, $C_{6h}$, or $S_{6}$ does not give rise to additional terms in the Hamiltonian and a result, all symmetry groups with inversion symmetry protect the structure of the $T$-invariant band inversion.

\subsubsection{Generalized Kane-Mele model}

We conclude this section by discussing the time-reversal invariant generalization of the honeycomb lattice model defined in Eq.~\eqref{eq:h-honeycomb}. As discussed, the honeycomb lattice model can be viewed as the Haldane model with third-nearest neighbor hopping across the hexagon. This immediately suggests that a time-reversal invariant version is obtained by replacing the Haldane term with the Kane-Mele spin-orbit coupling term \cite{kane2005b}. The Hamiltonian then becomes [see Eq.~\eqref{eq:h-honeycomb}]
\be
\mathcal H_\bk  = (\phi_\bk - t'\phi'_\bk)\tau_+  +  (\phi^*_\bk - t'\phi'^*_\bk)\tau_-  + t_{\text{soc}} \lambda^{f_1}_\bk \tau_z\sigma_z.  \label{eq:h-honeycomb-TRI}
\ee
The structure of this Hamiltonian is symmetry-protected as long as the symmetry group of the systems is $D_{6h}$ or $C_{6h}$.

\section{Discussion and conclusion \label{sec:discussion}}

In this work we have studied higher angular momentum band inversions in two dimensions. Owing to the non-vanishing density of states, these higher angular momentum band inversions provide a promising venue for realizing many-body generalizations of the topological phase transitions known from free fermion systems. To achieve such realizations two main directions for future research can be distinguished: the identification of materials which host higher angular momentum band inversions and a further investigation of the effect of interactions. 

The construction of lattice models presented in Sec.~\ref{sec:models} is a first and important step towards the identification of materials which exhibit rotation symmetry protected band inversions. In particular, we have shown that three different types of candidate systems can be identified, which differ in the nature of the microscopic degrees of freedom. Clearly, one possibility to realize an inversion of bands with relative angular momentum $m$ is to consider materials with local atomic orbital degrees of freedom, such as $d$- or $f$-wave states. As evidenced by an extended Haldane model on the honeycomb lattice, a second possibility relies on structures within a primitive unit cell, where wave functions transforming as higher angular momenta are formed by linear combinations of states at different sites. A third route relies on spin angular momentum states in strongly spin-orbit coupled systems. 

These systems provide potential avenues towards the experimental realization of many-body band inversions. It is important to note that realizing many-body Chern band transitions requires broken time-reversal symmetry, which may be challenging to realize other than by proximity to magnetic systems or in the presence of external magnetic fields. In this sense, a fruitful approach could target the time-reversal invariant generalizations discussed in Sec.~\ref{sec:T-invariant}, which rely on the presence of particular spin-orbit couplings.

A further direction for the future is the more detailed theoretical study of the effect of electronic interactions on the nature of the band inversion, focused in particular on the relative stability of the excitonic insulator phase and the fractional quantum Hall-type liquid states. In this regard the lattice models introduced in this work provide the basis for numerical studies. 

We conclude by pointing out that our work gives rise to an interesting application in three dimensions, specifically in the context of topological semimetals. Starting from the lattice models for higher angular momentum band inversions, one may introduce a third dimension and obtain a minimal model for a Weyl semimetal or a Dirac semimetal. More specifically, in case of the chiral models obtains lattice models for Weyl semimetals with higher monopole charge~\cite{fang2012b}. In fact, the square lattice model of Eq. \eqref{eq:hkC4} reproduces the double Weyl semimetal model introduced in Ref. \onlinecite{hassan2016}. The triangular lattice models can be promoted to lattice models for Weyl semimetals with monopole charge $C=3$. The time-reversal invariant generalizations introduced in Sec.~\ref{sec:T-invariant} can be extended with a coupling in the third dimension to produce simple lattice models for (band-inversion induced) Dirac semimetals. In this way the $T$-invariant model of Eq.~\eqref{eq:hk-D6h} gives rise to Dirac points with quadratic dispersion \cite{yang2014}.

\section*{Acknowledgements}

We thank Gene Mele, Stefanos Kourtis, and Michael Zaletel for helpful discussions. This work was supported by a Simons Investigator grant from the Simons Foundation and by National Science Foundation Grant DMR-1120901.

\appendix

\section{Ground state of Hamiltonian \eqref{eq:H} \label{app:pairing}}

The Hamiltonian $h_\bk$ of Eq. \eqref{eq:H} is diagonalized with the help of the unitary matrix $U_\bk$, which contains the eigenvectors as its columns, and one has
\be
U^\dagger_\bk  h_\bk U_\bk=  \begin{pmatrix} E_\bk & \\  & - E_\bk \end{pmatrix}, \quad  U_\bk=  \begin{pmatrix}   -u^*_\bk &v_{\bk} \\ v^*_\bk &  u_{\bk}  \end{pmatrix} \label{eq:Hdiag}
\ee
where $E_\bk = \sqrt{\varepsilon^2_\bk+|\Delta_\bk|^2}$ is the energy. The matrix $U_\bk$ must satisfy $U^\dagger_\bk U_\bk=1$, which implies $|u_\bk|^2+|v_\bk|^2=1$. The ratio of $u_\bk$ and $v_\bk$ is independent of the $U(1)$ phase degree of freedom associated with the eigenvectors and is given by
\be
v_\bk/u_\bk=  -(E_\bk-\xi_\bk)/ \Delta^*_\bk . \label{eq:ukvk}
\ee
We define normal mode operators $\gamma_{\bk e}$ and $\gamma_{\bk h}$ corresponding to the energy eigenvalues $\pm E_\bk$ as
\be
\gamma_\bk= \begin{pmatrix} \gamma_{\bk e} \\  \gamma_{\bk h} \end{pmatrix}=  U^\dagger \psi_\bk . \label{eq:chiChern}
\ee
The normal mode operators for the negative energy states are given by 
\be
  \gamma^\dagger_{\bk h} = v_{\bk} c^\dagger_{\bk e}+ u_{\bk} c^\dagger_{\bk h}  . \label{eq:gamma_h}
\ee
The mean-field ground state $\ket{\text{GS}}$ is given by filling all the negative energy states, i.e., $\ket{\text{GS}} = \prod_\bk \gamma^\dagger_{\bk h}\ket{0} $. Substituting Eq. \eqref{eq:gamma_h} and using the identity $c_{\bk h}c^\dagger_{\bk h}\ket{0} = \ket{0}$, the ground state can be written in the following form
\be
\ket{\text{GS}} = \prod_\bk (u_{\bk} + v_{\bk} c^\dagger_{\bk e} c_{\bk h})\ket{\Omega}  . \label{eq:GS}
\ee
Here $\ket{\Omega}$ defines a vacuum state obtained by filling all valence band states: $\ket{\Omega} = \prod_\bk c^\dagger_{\bk h}\ket{0}$. Since $ c_{\bk h}$ creates holes in the vacuum defined by $\ket{\Omega}$, it is natural to perform a particle-hole transformation on the hole operators given by
\be
 c_{\bk h}  \to   c^\dagger_{-\bk h}, \quad   \gamma_{\bk h}  \to   \gamma^\dagger_{-\bk h} . \label{eq:ehdef}
\ee
After particle-hole transformation the normal mode annihilation operators take the form
\beq
\gamma_{\bk e} & = &  v_{\bk} c^\dagger_{-\bk h}-u_\bk  c_{\bk e}  \\
\gamma_{-\bk h} & = &  v_{\bk} c^\dagger_{\bk e}+ u_{\bk} c_{-\bk h}
\eeq
and in full analogy with BCS theory one obtains the ground state by determining the state which is annihilated by all such normal mode operators. A state which clearly has this property is $ \prod_\bk \gamma_{\bk e}\gamma_{-\bk h}\ket{\Omega}$ and one thus finds the ground state as
\be
\ket{\text{GS}} = \prod_\bk (u_{\bk} + v_{\bk} c^\dagger_{\bk e} c^\dagger_{-\bk h})\ket{\Omega}, \label{eq:GS2}
\ee
which is precisely \eqref{eq:GS} with $c_{\bk h}  \to   c^\dagger_{-\bk h}$.

\section{Excitonic insulator mean field theory \label{app:mftheory}}

For the purpose of a mean field analysis it is useful to express  interacting Hamiltonian $H_U$ of Eq. \eqref{eq:H-U-k} in a form which can decoupled. To this end we rewrite the interacting Hamiltonian as 
\be
H_U = -\frac{U}{4 N} \sum_{\bk\bk'} ( \varphi^\dagger_{\bk} \tau_x\varphi_\bk )(\varphi^\dagger_{\bk'} \tau_x\varphi_{\bk'}),  \label{appeq:H-U-k-2}
\ee
with $\varphi_\bk$ as defined in Eq. \eqref{eq:basisC6}. Here $N$ is the system size. To perform the mean field decoupling of the interaction, we write the action of interacting system as $S=S_0 +S_U$ with $S_0 = \int_0^\beta d\tau \sum_\bk   \varphi^\dagger_{\bk}(\partial_\tau + h_\bk)\varphi_\bk$
and $S_U= \int_0^\beta d\tau H_U$. The interacting part of the action is decoupled in terms of the field $\Delta_0$ as $\exp(-S_U) = \int\mathcal{D}\Delta_0 \exp(-S_U'[\Delta_0]) $, where $S_U'[\Delta_0] $ is now bilinear in the fermions and given by $S_U'[\Delta_0]  = \int_0^\beta d\tau H_U'[\Delta_0] $ with
\be
H'[\Delta_0] = 2\Delta_0  \sum_{\bk}\varphi^\dagger_{\bk} \tau_x\varphi_\bk + 4N\Delta^2_0/U.  \label{appeq:H'}
\ee
For the subsequent analysis it is convenient to redefine the mean field as $2\Delta_0 \to \Delta_0$.

\subsection{Mean field solution   \label{app:MF-solution}}

Integrating out the fermions one obtains the free energy as a functional of $\Delta_0$; the saddle-point of this free energy defines the mean field self-consistency equation, which is given by
\be
\frac{\delta F}{\delta \Delta_0}  = 0 \; \Rightarrow \; \Delta_0 = -\frac{U}{2N}\sum_{\bk} \langle   \varphi^\dagger_\bk \tau_x\varphi_\bk \rangle.  \label{appeq:stationary}
\ee
The expectation value is defined with respect to the ground state of the mean field Hamiltonian 
\be
h^{\text{MF}}_\bk=  \begin{pmatrix}  \varepsilon_\bk & \Sigma_\bk  \\ \Sigma^*_\bk  & - \varepsilon_\bk \end{pmatrix} ,\quad \Sigma_\bk= \Delta_0 + \Delta_\bk, \label{eq:h-k-MF}
\ee
where $\Delta_\bk = \Delta_3(i \lambda^{f_1}_\bk+\lambda^{f_2}_\bk)$. Here we have taken $\Delta_1=\Delta_2$ in Eq. \eqref{eq:hkC6} and redefined it as $\Delta_{m=3}$. The energies of the mean field Hamiltonian are given by $ \pm E_\bk = \pm \sqrt{\varepsilon^2_\bk + |\Sigma_\bk|^2}$ and the matrix $U_\bk$ which diagonalizes the mean field Hamiltonian is given by
\be
U_\bk=  \frac{1}{\sqrt{2E_\bk(E_\bk-\varepsilon_\bk)}} \begin{pmatrix}   -\Sigma^*_\bk & \varepsilon_\bk-E_\bk \\ \varepsilon_\bk-E_\bk & \Sigma_\bk \end{pmatrix}. \label{appeq:U-k}
\ee
Substituting this into the self-consistency condition \eqref{appeq:stationary} one finds
\be
\Delta_0 =  \frac{U}{2N} \sum_{\bk} \frac{\text{Re}[\Sigma_\bk] }{E_\bk} [ f(-E_\bk)-f(E_\bk)].  \label{appeq:solution}
\ee
where $f(\varepsilon) = (1+e^{\beta \varepsilon})^{-1}$ is the Fermi function. At zero temperature the self-consistency condition reduces to $\Delta_0 =  \frac{U}{2N} \sum_{\bk}\text{Re}[\Sigma_\bk] /E_\bk$.

\subsection{Free energy \label{app:F}}

The free energy itself can be directly evaluated and at finite temperature $T$
\be
F[\Delta_0]  = -\frac{2}{\beta }\sum_{\bk} \ln \cosh\left(\frac{\beta E_\bk}{2}\right) + \frac{N}{U}\Delta^2_0 ,  \label{eq:F-T}
\ee
where we have ignored the constant contribution $-(N/\beta)\ln 4$. By taking the derivative with respect to $\Delta_0$ and setting it equal to zero one recovers the saddle-point equation \eqref{appeq:solution}.

At zero temperature the free energy takes the simple form 
\be
F[\Delta_0]  = -\sum_{\bk} E_\bk + N\Delta^2_0/U.  \label{eq:F-T=0}
\ee

\subsection{Expansion of free energy in $\Delta_0$ \label{app:F}}

To study the phase transition to the rotation symmetry broken state one may expand the free energy powers of the order parameter $\Delta_0$ to obtain a simple Landau theory for the transition. At zero temperature, the free energy \eqref{eq:F-T=0} can be expanded as 
\be
F[\Delta_0]/N  = \left(U^{-1}-c_2\right) \Delta^2_0 +c_4 \Delta^4_0,  \label{eq:F}
\ee
where the expansion coefficients are given by
\beq
c_2 & = &  \frac{1}{2N}\sum_\bk \left(\frac{1}{E_\bk} - \frac{(\text{Re}\Delta_\bk )^2}{E^3_\bk} \right), \\
c_4 & = & \frac{1}{8N}\sum_\bk  \left(\frac{1}{E^3_\bk} - 6\frac{(\text{Re}\Delta_\bk )^2}{E^5_\bk}+ 5\frac{(\text{Re}\Delta_\bk )^4}{E^7_\bk} \right). 
\eeq
Note that since we are expanding around $\Delta_0=0$, in these expressions the energy $E_\bk$ is evaluated at $\Delta_0$, i.e., $E_\bk = \sqrt{\varepsilon^2_\bk + |\Delta_\bk|^2}$.

\end{document}